\documentclass[prd,twocolumn,reprint,preprintnumbers,nofootinbib,superscriptaddress,longbibliography]{revtex4-1}

\usepackage{slashed,color}
\usepackage{dcolumn}

\usepackage{bm}
\usepackage{amsmath} 
\allowdisplaybreaks
\usepackage{amssymb}  
\usepackage{enumerate}
\usepackage{multirow}
\usepackage{braket}
\usepackage[toc,page]{appendix}
\usepackage{url}
\usepackage{graphicx}
\usepackage{hyperref}
\hypersetup{colorlinks=true}
\usepackage{latexsym}
\usepackage{wasysym}
\usepackage{dhucs-enumerate}
\usepackage[most]{tcolorbox}

\usepackage{setspace}
\usepackage{mathrsfs,amssymb}  
\usepackage{cancel}
\usepackage[normalem]{ulem}
\usepackage{cleveref}
\usepackage{tikz}
\usetikzlibrary{decorations.markings}
\usetikzlibrary{positioning}
\usetikzlibrary{shapes}
\usetikzlibrary{calc}
\usepackage{textcomp}
\usepackage{lipsum}
\usepackage{newtxtext, newtxmath}
\usepackage[export]{adjustbox}
\usepackage{tabularray}
\UseTblrLibrary{booktabs}
\usepackage{float}

\let\ket\relax

\usepackage[braket, qm]{qcircuit}

\usepackage{hhline}
\usepackage{soul}

\begin{document}

\title{
Quantum Integration Networks for Efficient Monte Carlo in High-Energy Physics}

\author{Heechan Yi}
\email{lhc320@yonsei.ac.kr}
\affiliation{Department of Physics, Yonsei University, Seoul 03722, Republic of Korea}

\author{Kayoung Ban}
\email{kban@kias.re.kr}
\affiliation{School of Physics, Korea Institute for Advanced Study, Seoul 02455, Republic of Korea}

\author{Myeonghun Park}
\email{parc.seoultech@seoultech.ac.kr}
\affiliation{School of Natural Sciences, Seoultech, Seoul 01811, Republic of Korea}
\affiliation{Institute of Convergent Fundamental Studies, Seoultech, Seoul 01811, South Korea}
\affiliation{Institute for Basic Science, Daejeon 34126, South Korea}%

\author{Kyoungchul Kong}
\email{kckong@ku.edu}
\affiliation{Department of Physics and Astronomy, University of Kansas, Lawrence, KS 66045, USA}

\begin{abstract}

\begin{spacing}{1.2}
Monte Carlo methods play a central role in particle physics, where they are indispensable for simulating scattering processes, modeling detector responses, and performing multi-dimensional integrals. However, traditional Monte Carlo methods often suffer from slow convergence and insufficient precision, particularly for functions with singular features such as rapidly varying regions or narrow peaks.
Quantum circuits provide a promising alternative: compared to conventional neural networks, they can achieve rich expressivity with fewer parameters, and the parameter-shift rule provides an exact analytic form for circuit gradients, ensuring precise optimization. Motivated by these advantages, we investigate how sampling strategies and loss functions affect integration efficiency within the \textbf{Quantum Integration Network} (QuInt-Net). We compare adaptive and non-adaptive sampling approaches and examine the impact of different loss functions on accuracy and convergence. Furthermore, we explore three quantum circuit architectures for numerical integration: the data re-uploading model, the quantum signal processing protocol, and deterministic quantum computation with one qubit. The results provide new insights into optimizing QuInt-Nets for applications in high energy physics.
\end{spacing}

\end{abstract}


\maketitle


\newpage


\section{Introduction}\label{sec:intro}

In high energy physics, numerical integration is an essential component of numerous calculations, particularly those related to the determination of physical quantities such as scattering amplitudes and cross-sections \cite{metropolis1949journal, caflisch1998monte,zhong2023efficient,lepage1978new,lepage2021adaptive,Shyamsundar:2023jtz}. 
Indeed, the cross-section ($\sigma$) for any given scattering process is defined by a multi-dimensional integral over the final state phase space.
To extract information on a specific variable 
$X(p_1,\ldots,p_n)$, one needs to evaluate the corresponding differential distribution,
\begin{align} 
\frac{d\sigma}{d X} = &\frac{1}{2s}\int\prod_{i=1}^{n} \frac{d^3 \vec p_i}{(2\pi)^3 2E_i} \delta\left(k_1+k_2-\sum_{i=1}^n p_i\right) \nonumber \\
&\times |M_{fi}|^2 \theta_{\rm cut}(p_1,\cdots, p_n)\delta (X-X(p_1,\cdots,p_n)) \, ,
\end{align} 
where $p_i = (E_i, \vec p_i)$ denotes the energy and momentum of particle $i$ ($i=1,\cdots , n$) in the final state. 
The first Dirac delta function implies the energy-momentum conservation, and the second ensures the constraint on the $X$ variable. 
The $\theta_{\rm cut}$ denotes the reduction of phase space due to acceptance or kinematic cuts.  
The accurate evaluation of the integrals is essential for collider physics, yet presents substantial computational challenges. 
The complexity of multidimensional integration grows rapidly with the number of final state particles, and the integrand often develops complicated structures, such as resonant propagators or kinematic thresholds.
These features lead to resonance peaks and singular regions, which make numerical integration inefficient or even unstable.
As a result, reliable predictions typically require sophisticated integration methods together with tailored mappings of the integration variables to control singular behavior and enhance convergence.

Monte Carlo (MC) integration is widely used for its scalability to high-dimensional spaces. It also allows for binning of partial results, facilitating the extraction of differential distributions across various integration variables.
However, MC integration suffers from slow convergence characterized by a convergence rate of $O(N^{-1/2})$, requiring a large number of samples ($N$) for accurate estimation (see \cite{Shyamsundar:2023jtz} and the references therein.). 
Moreover, it struggles with localized features in the integrand, such as narrow resonance peaks, where simple random sampling often fails to resolve the relevant features of the integrand.
The problems in numerical integration have led to the development of more effective methods, ranging from early Monte Carlo approaches \cite{metropolis1949journal, caflisch1998monte} to recent techniques specifically designed for high energy physics \cite{zhong2023efficient}.

The VEGAS algorithm~\cite{lepage1978new, lepage2021adaptive}, widely used in particle physics, employs adaptive importance sampling to reduce variance by iteratively refining the sampling distribution.
In recent years, several research groups have worked to improve the algorithm without modifying its fundamental sampling strategy.
For instance, implementations on modern hardware, such as GPUs and tensor based frameworks, have demonstrated significant speed ups~\cite{carrazza2020vegasflow, gomez2021torchquad}.
More recently, machine learning approaches have been explored to model complex sampling distributions using neural networks, leading to more efficient integration in high dimensional phase spaces \cite{muller2019neural, bothmann2020exploring, hammad2023exploration, ban2024lestrat}.
%
Furthermore, normalizing flows and invertible neural networks have been applied to phase space integration~\cite{gao2020event, heimel2023madnis, ernst2023normalizing}.
Recent studies introduced several alternatives to Monte Carlo integration, transforming the sampling problem into a function approximation task. Some of the proposals have been extended to the quantum machine learning (QML) domain~\cite{cruz2024multi,schuld2015introduction, biamonte2017quantum, mitarai2018quantum, perez2020data, chen2020variational}. 
The hybrid quantum-classical approach \cite{cruz2024multi} is further developed to compute Feynman loop integrals with Quantum Fourier Iterative Amplitude Estimation~\cite{martinez2024loop, de2024quantum}.

In this study, we introduce QuInt-Net (Quantum Integration Network) as a proposed framework for quantum circuit based function integration, and refer to it throughout this work.
We investigate how training quantum circuits can improve integration accuracy for functions with prominent features and facilitate function approximation. 
Building on the work of \cite{cruz2024multi}, we adapt the learning strategy to improve the estimation of complex integrals, extending it beyond smooth integrands to explicitly address functions with singular structures such as discontinuities or resonance peaks. 
We investigate how learning strategies, particularly data quality and loss function design, influence the accuracy of function approximation.
Specifically, we adapt Importance Sampling (IS)~\cite{tokdar2010importance} and Hamiltonian Monte Carlo (HMC)~\cite{neal2011mcmc, betancourt2017conceptual, hoffman2014no} to target regions of high functional variance, and construct composite loss functions to better capture singular structures such as peaks or discontinuities. 

Quantum circuits have advantages for calculating the gradient of the circuit output using the hardware itself.
In particular, the Parameter Shift Rule (PSR) enables efficient gradient evaluation within the same circuit architecture for both the function and its derivative \cite{schuld2019evaluating, Mari_2021, wierichs2022general}.
We investigate the general variational quantum circuit model for Quantum Neural Network (QNN), which employs the data re-uploading method~\cite{perez2020data}.
Furthermore, we explore the application to alternative quantum circuit Ans\"atze, including the Quantum Signal Processing (QSP) protocol \cite{Martyn_2021, motlagh2024generalized} and Deterministic Quantum Computation with One qubit (DQC1) \cite{knill1998power, kim2025expressivity}. 
Since the data-reuploading QNN provides the most reliable performance according to our evaluation metrics (see Section~\ref{sec:metric}), we present its results as the main findings in this paper.
The other results using QSP and DQC1 are introduced in the Appendix~\ref{appendix:result}.

This paper is structured as follows. 
Section~\ref{sec:Methodology} introduces the outlines of the methodology, including data sampling strategies, and loss functions design used for model training. 
Section~\ref{sec:Result} demonstrates the results and performance of each proposed approach in representative toy examples and high energy physics applications. 
Finally, Section~\ref{sec:Conclusion} summarizes the key findings, discusses limitations, and outlines potential directions for future research.
Reference code with examples is publicly available on the web\footnote{Code available on \url{https://github.com/HeechanYi/QML_Integral}}.

\section{Methodology}\label{sec:Methodology}

We introduce QuInt-Net, a framework for computing function integrals. 
We detail the sampling strategy used to construct the training set, and describe the design of the corresponding loss function. 

Here, we adopt variational quantum circuits (VQCs) with data re-uploading quantum neural networks (QNN) to implement QuInt-Net.
For more detail, the circuit structures for QNN and alternative models (QSP and DQC1) are provided in Appendix~\ref{appendix:result}, along with their corresponding experimental results. 

\subsection{Quantum Integration Network (QuInt-Net)}

\begin{figure*}[t]
    \includegraphics[width = 0.95\textwidth]{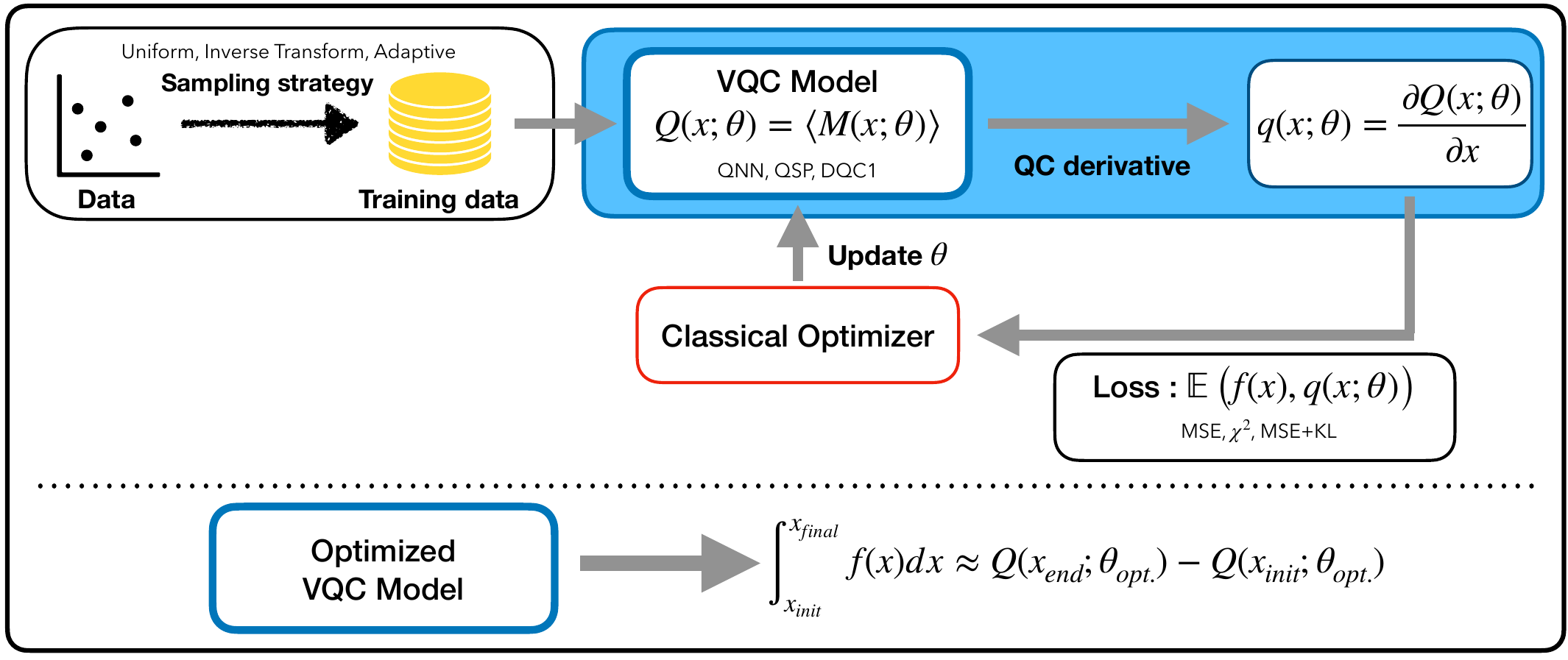}
    \caption{Schematic illustration of the variational quantum circuit (VQC) workflow for numerical integration. Training data are generated using various sampling strategies, including uniform, importance, and adaptive sampling. The VQC model $Q(x;\theta)$, implemented with QNN, QSP, or DQC1 architectures, is trained to approximate both $f(x)$ and its derivative $q(x)=\partial_x Q(x;\theta)$. A classical optimizer minimizes a loss function such as MSE, $\chi^2$, Log-Cosh, or KL-augmented MSE. The trained VQC model can then be used to calculate the integral of the target function via the optimized parameters.}
    \label{fig:VQCdiagram}
\end{figure*}

We use VQCs, which are the quantum analogs of classical parametric models ~\cite{schuld2015introduction, biamonte2017quantum}.
VQCs are composed of parameterized quantum gates whose trainable parameters, $\theta$, control the evolution of a quantum state.
The unitary operator $U(\theta)$ transforms an initial state $|\psi_i\rangle$ into a final state $|\psi_f\rangle$.
We then measure an observable $\hat{O}$ on the final state,
\begin{equation}\label{eqn:(2)}
    Q_{\hat{O}}(\theta) = \braket{\psi_i|U^{\dagger}(\theta)\hat{O}U(\theta)|\psi_i},
\end{equation}
and we follow the classical machine learning strategy, designating the (one dimensional) input as $x$  and the (one dimensional) target function as $y=f(x)$ where $f$ is the target integrand in this study.
We denote the circuit output as $Q(x;\theta)$, which is trained to approximate the target function $f(x)$.

There are several methods for embedding data into a quantum circuit system \cite{lloyd2020quantum, perez2020data, schuld2018supervised}.
We use the data re-uploading approach in this study, which embeds the input data $x$ directly in parameterized gates, as outlined in Ref.~\cite{cruz2024multi}.
A key advantage of this approach is that it requires no additional state preparation.
The initial state is chosen to be the ground state, $|\psi_i\rangle = |0\rangle$.
The circuit output thus becomes an explicit function of $x$, changing Eq.~\eqref{eqn:(2)} to:
\begin{equation}
    Q(x; \theta) = \braket{0|U^{\dagger}(x;\theta)\hat{O}U(x;\theta)|0}.
\end{equation}

To approximate the integrand $f(x)$, we use the derivative of the circuit output with respect to its input, $\partial_x Q(x;\theta)$, which can be computed efficiently using the Parameter Shift Rule (PSR). 
The PSR provides exact gradient estimates, but only when the generator of a parameterized gate has exactly two distinct eigenvalues~\cite{schuld2019evaluating, Mari_2021, wierichs2022general}. 
However, since the VQC model involves many parameterized gates, PSR-based gradient computation becomes slow.
Moreover, for alternative circuit models such as QSP and DQC1, the use of two-qubit gates introduces additional complexity for PSR-based calculations~\cite{anselmetti2021local}. 
However, as we adopt the simulation based computation, we use classical backpropagation to compute gradients, which offers greater efficiency for large circuits.

The goal of training is to optimize the circuit parameters $\theta$ such that the derivative of the VQC output, $q(x;\theta) \equiv \partial_x Q(x; \theta)$, approximates the target function $f(x)$ within the integral limits $(x_{\rm init}, x_{\rm final})$.
We generate a training dataset of $N_{\mathrm{train}}$ points, $\{x_i, f(x_i)\}$, within the integration domain $[x_{\mathrm{init}}, x_{\mathrm{final}}]$. 
Then, we minimize a loss function ($\mathbb{E}$) that quantifies the deviation between our model's derivative and the target function across the training data as
\begin{equation}
    \mathbb{E}[f(x_i), q(x_i;\theta)].
\end{equation}
We minimize this loss using the Adam optimizer~\cite{kingma2014adam} to find the optimal parameters, $\theta_{\mathrm{opt}}$.
So after the optimization, the resulting model, $Q(x; \theta_{\mathrm{opt}})$, serves as the learned antiderivative of $f(x)$. 
The definite integral is then computed directly as
\begin{equation}
    I = \int_{x_{\mathrm{init}}}^{x_{\mathrm{final}}} f(x) dx \approx Q(x_{\mathrm{final}}; \theta_{\mathrm{opt}}) - Q(x_{\mathrm{init}}; \theta_{\mathrm{opt}}).
\end{equation}
A key advantage of the method is that the trained circuit can compute the integral over any subinterval within the training domain without retraining. 
The entire procedure is outlined in Fig.~\ref{fig:VQCdiagram}.

\subsection{Data Sampling}

Given that training data quality is a key factor in determining model performance, we construct a training dataset using three sampling strategies: uniform sampling, Importance Sampling (IS), and Hamiltonian Monte Carlo (HMC). 
In this section, we explain how we use IS and HMC to make the data set for training.

\subsubsection{Importance Sampling (IS)}

We use Importance Sampling (IS), a variance reduction technique that distributes samples to the integrand's most important regions~\cite{tokdar2010importance}.
A lower variance estimate and quicker convergence are the results of correcting for the deliberate bias introduced by the proposal distribution $q(x)$ using the re-weighting factor $f(x)/q(x)$.
Since the features frequently dominate the integral's value, it focuses on regions where the integrand $f(x)$ displays singular structures, such as resonances and discontinuities.

To elaborate, we first create data points from a uniform distribution, $q_{\mathrm{uniform}}(x)$, over the domain $[x_{\mathrm{init}}, x_{\mathrm{final}}]$.
Then, we create a target distribution, $q_{\mathrm{target}}(x) \propto (f'(x))^2$, proportional to the square of the integrand's derivative.
Each candidate point  $x_i$  is then given a weight $w_i \propto (f'(x_i))^2$.
The reweighted data pool is sampled with replacement to produce the final training data set.
By creating a dataset focused on the integrand's difficult characteristics, this process makes it possible for the VQC to be trained more successfully.

\subsubsection{Hamiltonian Monte Carlo (HMC)}

\begin{figure*}[ht!]
    \includegraphics[width = 0.9\textwidth]{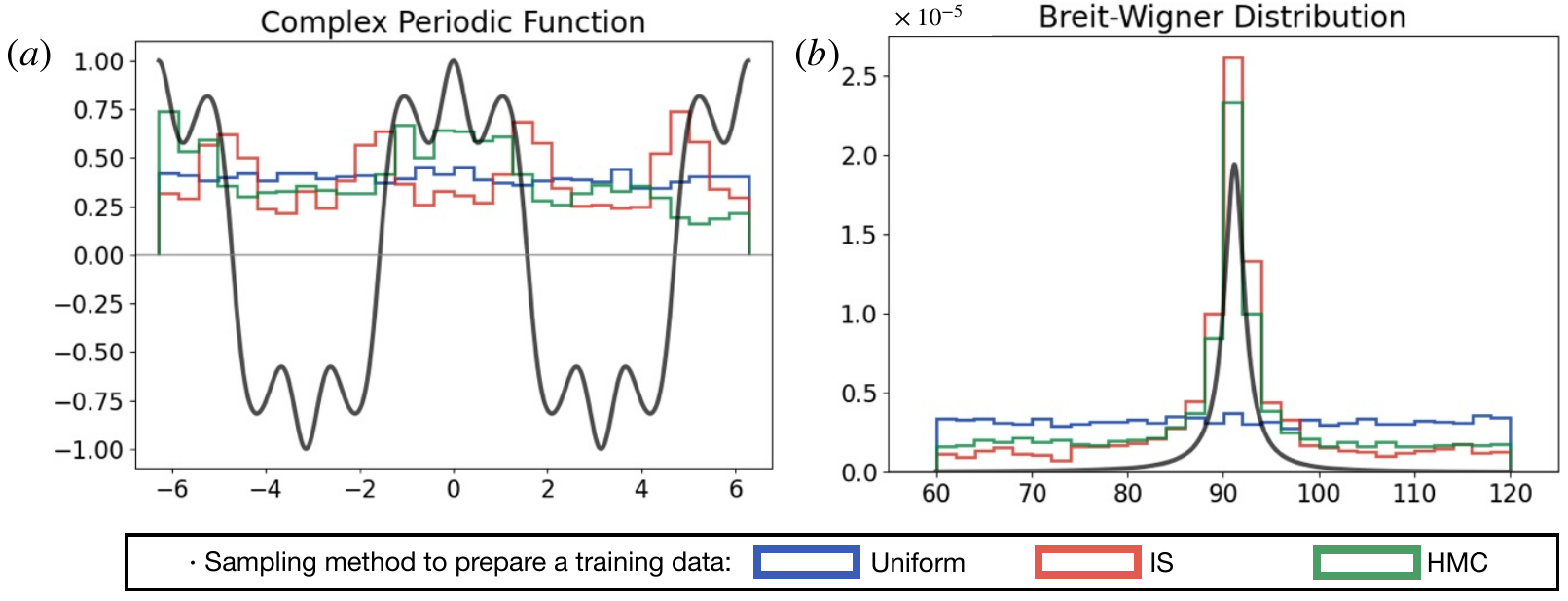}
    \caption{
    Sample distributions obtained from three different sampling strategies: uniform (blue), importance sampling (IS, red), and Hamiltonian Monte Carlo (HMC, green). The target function is shown with black solid lines.
    Figure in (a) shows results for the Complex Periodic Function (CPF), a highly oscillatory integrand.
    Figure in (b) shows the Breit–Wigner distribution, where both IS and HMC successfully concentrate samples near the resonance peak, demonstrating their effectiveness in capturing localized features.
    Each functions are introduced in Section~\ref{sec:CPF} and Sec~\ref{sec:BW}.}
    \label{fig: Sampling}
\end{figure*}
As an alternative, we adapt Hamiltonian Monte Carlo (HMC), a Markov Chain Monte Carlo (MCMC) technique that is especially useful for investigating intricate probability distributions~\cite{neal2011mcmc, betancourt2017conceptual}.
It allows for effective, large scale movements through the sample space because it is based on the Hamiltonian dynamics for a hypothetical particle on a potential energy surface.
We define the potential energy from the target integrand $f(x)$ for our numerical integration task as $U(x) = -\log|f(x)|$. 
To ensure numerical stability, a small regularization constant is added to $|f(x)|$.

The data point travels through the Hamiltonian system $\mathcal{H}(x, p)$, defined as:
\begin{equation}
    \mathcal{H}(x, p) = U(x) + K(p),
\end{equation}
where $x$ is the position, $p$ is an auxiliary momentum variable, which corresponds to a standard kinetic energy term $K(p)=p^2/2$.
Starting from a random position $x_t$, we choose the momentum with the Gaussian distribution, $p_t \sim \mathcal{N}(0, \mathbf{I})$.
The trajectory of this system in phase space evolves according to Hamilton's equations of motion:
\begin{align}
    \frac{dx}{dt} &= \frac{\partial \mathcal{H}}{\partial p} = \frac{\partial K}{\partial p}, \\
    \frac{dp}{dt} &= -\frac{\partial \mathcal{H}}{\partial x} = -\frac{\partial U}{\partial x}.
\end{align}
We numerically propagate these equations for $n=20$ steps with step size $\epsilon = 0.1$ using a leapfrog integrator.
The required potential gradient, $\nabla U(x)$, is computed via a central finite-difference method.
The final state of the trajectory is then accepted or rejected using a Metropolis-Hastings criterion~\cite{metropolis1953equation, hastings1970monte, robert2009metropolis}, and we reject any moves outside the domain $[x_{\mathrm{init}}, x_{\mathrm{final}}]$.

To avoid a too biased distribution, we combine the samples generated by HMC and from uniform sampling to train our model.
Then, the dataset provides a balance between the global and singular structure of the integrand function.
Finally, Fig.~\ref{fig: Sampling} demonstrates the data distribution of our adaptive strategies, showing how both IS and HMC successfully concentrate samples on the key features of representative functions in contrast to uniform sampling.

\subsection{Loss Function Design}

The selection of the loss function is crucial as it directly influences the updating of parameters.
In this study, we test several loss functions to assess their effect on training VQC model for functions exhibiting singular structures.
We denote the derivative of the model output as $q(x; \theta) = \partial_x Q(x; \theta)$, and the target integrand as $f(x)$.

In general, we use the basic Mean Squared Error (MSE), defined as:
\begin{equation}
    \mathbb{E}_{\mathrm{MSE}} = \frac{1}{N_{\mathrm{train}}} \sum_{i = 1}^{N_{\mathrm{train}}}  \left[ q(x_i|\theta) - f(x_i) \right]^2.
    \label{eqn:MSE}
\end{equation}
While MSE shows a stable efficiency, it often struggles when the integrand's singularities are dominated by the largest residuals.

To address the limitation and to improve sensitivity to lower-magnitude regions, we consider a weighted loss inspired by the $\chi^2$ statistic:
\begin{equation}
    \mathbb{E}_{\chi^2} = \frac{1}{N_{\mathrm{train}}} \sum_{i = 1}^{N_{\mathrm{train}}} \frac{ \left( q(x_i|\theta) - f(x_i) \right)^2 }{ |f(x_i)| + \epsilon }.
    \label{eqn:chi}
\end{equation}
The loss function re-weights squared residuals to ensure that distributional tails are fitted with comparable fidelity to central regions.
However, it can be numerically unstable if $f(x_i)$ is near zero (hence the regularization constant $\epsilon$) and may cause underfitting of sharp peaks by suppressing their large contributions.

As an alternative that is more robust to large errors, we examine the Log-Cosh loss, which smoothly interpolates between MSE and Mean Absolute Error:
\begin{equation}
    \mathbb{E}_{\mathrm{Log-Cosh}} = \frac{1}{N_{\mathrm{train}}}\sum_{i = 1}^{N_{\mathrm{train}}} \log\left(\cosh (q(x_i | \theta) - f(x_i))\right).
    \label{eqn:Log_Cosh}
\end{equation}
The Log-Cosh loss behaves linearly for large residuals, making it less sensitive to outliers than MSE, while remaining smooth and twice-differentiable for stable optimization.

We propose a composite loss function that combines MSE with the Kullback–Leibler (KL) divergence to move beyond point-wise errors and capture the overall shape of the integrand:
\begin{equation}
\begin{split}
    \mathbb{E}_{\mathrm{MSE+KL}} =\mathbb{E}_{\mathrm{MSE}} + \lambda \sum_{i = 1}^{N_{\mathrm{train}}} \mathrm{KL} \left( \sigma(\mathbf{f}) \, \| \, \sigma\left( \mathbf{q}_{\theta} \right) \right).
\end{split}
\label{eqn:MSE_KL}
\end{equation}
Target values and model predictions are denoted as $\mathbf{f} = \{f(x_i)\}$ and $\mathbf{q}_{\theta} = \{q(x_i; \theta)\}$ across a training batch as inputs to a softmax function, $\sigma$, here. 
The transform makes the target function and model output into probability distributions, allowing the KL divergence to measure the difference in their relative shapes. 
By minimizing both the direct error and the distributional difference, this loss encourages the model to learn not just the local values but also the global structure of the target function.

\section{Results}\label{sec:Result}

Having defined the training objective, we proceed to assess our approach using benchmark functions that embody distinct structural challenges.
It includes (i) a multi-scale oscillatory function (Complex Periodic Function), (ii) a discontinuous step function, and (iii) the Breit-Wigner distribution, which is relevant to resonance phenomena in high energy physics.
Each of the cases shows a distinct type of localized or singular structure, demanding high-fidelity modeling to ensure reliable integral estimation.

\subsection{Metrics for Model Performance Estimation}\label{sec:metric}

To estimate the model performance, we consider three evaluation metrics.
We quantify 
(i) the point-wise accuracy of the learning process, the approximation of the $q(x;\theta)$ to the target function $f(x)$, with the coefficient of determination, $R^2$ score;~\cite{steel1960principles, glantz1990primer, draper1998applied}
(ii) the global integral behavior of the model by the Wasserstein distance, $W_1$, to check the performance of the integral over the total training interval;~\cite{kantorovich1960mathematical, vaserstein1969markov, vallander1973calculation}
and (iii) the integration accuracy over three subintervals containing the singular features, which represents the challenging region for calculating the integral.

The $R^2$ is estimated as:
\begin{align}
     R^2 & = 1 - \frac{1}{N\sigma^2} \sum_{i=1}^N \left(f(x_i) - q(x_i; \theta) \right)^2
\end{align}
where $f(x_i)$ is the value of the target function and $q(x_i; \theta)$ is the model's predicted derivative at point $x_i$. 
$N$ is the number of test points in the total training range and $\sigma^2$ is the variance of the target values with the test points.
Then, the $R^2$ score estimates how well the model has been trained, and further, it provides the reliability of the model’s integration results.

The $W_1$ distance is defined as:
\begin{align}
     W_1 & = \sum_{m=1}^M |J_{\mathrm{true}}^{\mathrm{norm}}(x_m) - J_{\mathrm{pred}}^{\mathrm{norm}}(x_m)| \cdot \Delta x_m
\end{align}
where $\Delta x_m  = |x_k - x_{k-1}|$.

The interval is divided into $M=30$ equal subintervals. 
The $J^{\mathrm{norm}}(x_m)$ is defined by accumulating the integrals from the first to the $m$-th subinterval and then normalizing by the maximum cumulative value across the subintervals.
Consequently, $J(x_m)^{\mathrm{norm}}$ is calculated like:
\begin{align}
    J^{\mathrm{norm}}(x_m) & = \frac{\sum_{k=1}^m I(x_k)}{\max\limits_{1\leq n \leq M} \sum_{k=1}^n I(x_k)}, \nonumber 
\end{align}
where $x_k$ is the endpoint of the $k$-th subintervals and $I(x_k) = \int_{x_{k-1}}^{x_k} f(x')dx'$ is the definite integral over $k$-th subinterval.
It is calculated for both the target integral and predicted output, represented as $J_{\mathrm{true}}(x_m)$ and $J_{\mathrm{pred}}(x_m)$ respectively.
Finally, the $W_1$ distance is evaluated with the absolute difference between the two normalized cumulative functions, $J_{\mathrm{true}}(x)$ and $J_{\mathrm{pred}}(x)$.
The $W_1$ distance estimates the global integral behavior, and also the model performance in calculating the integral.

\begin{figure*}[t]
  \centering
  \begin{minipage}[t]{0.33\textwidth}
    \includegraphics[width=\linewidth,
                     height=0.28\textheight, keepaspectratio]
      {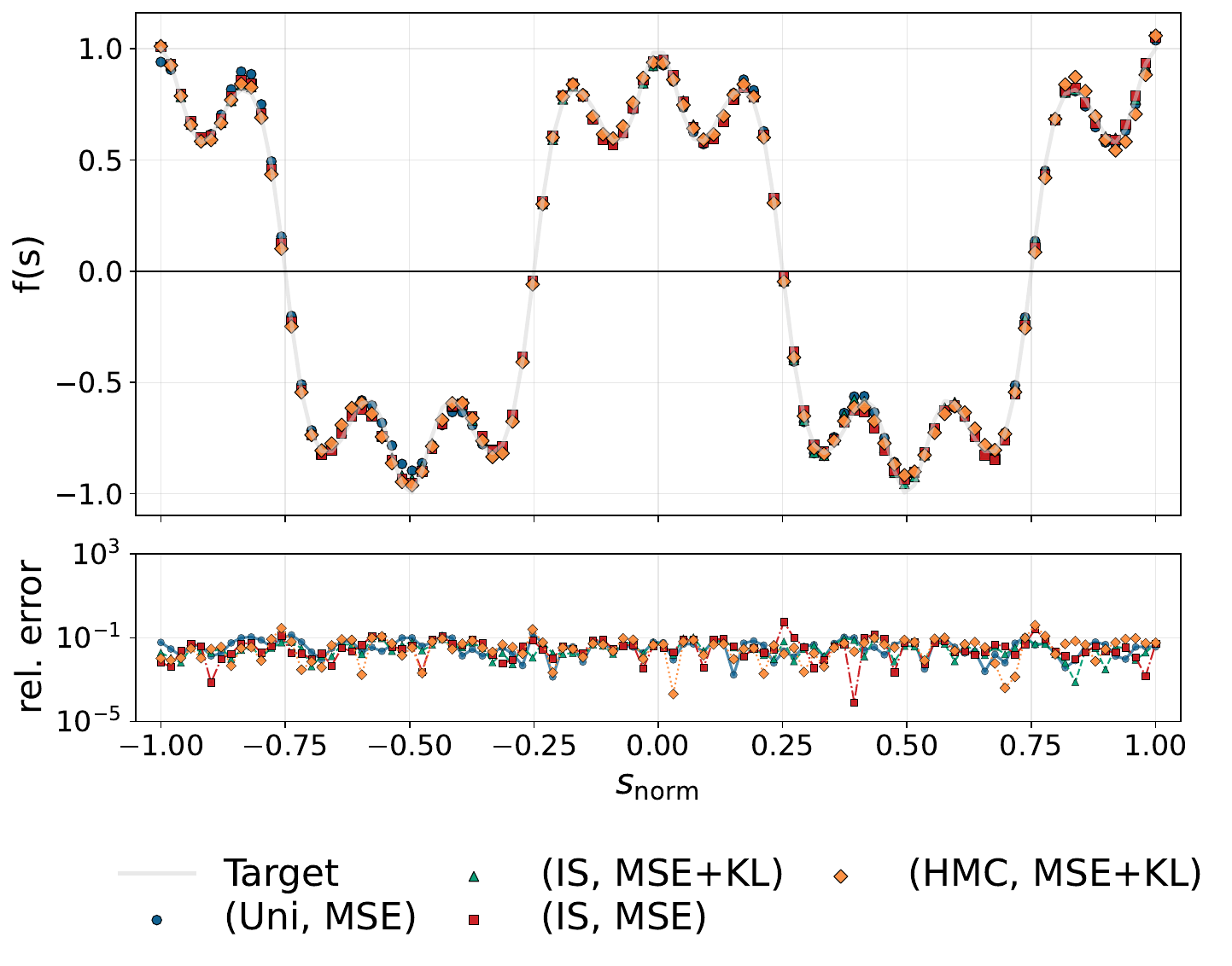}
    \par (a) Training result
  \end{minipage}
  \begin{minipage}[t]{0.33\textwidth}
    \includegraphics[width=\linewidth,
                     height=0.28\textheight, keepaspectratio]
      {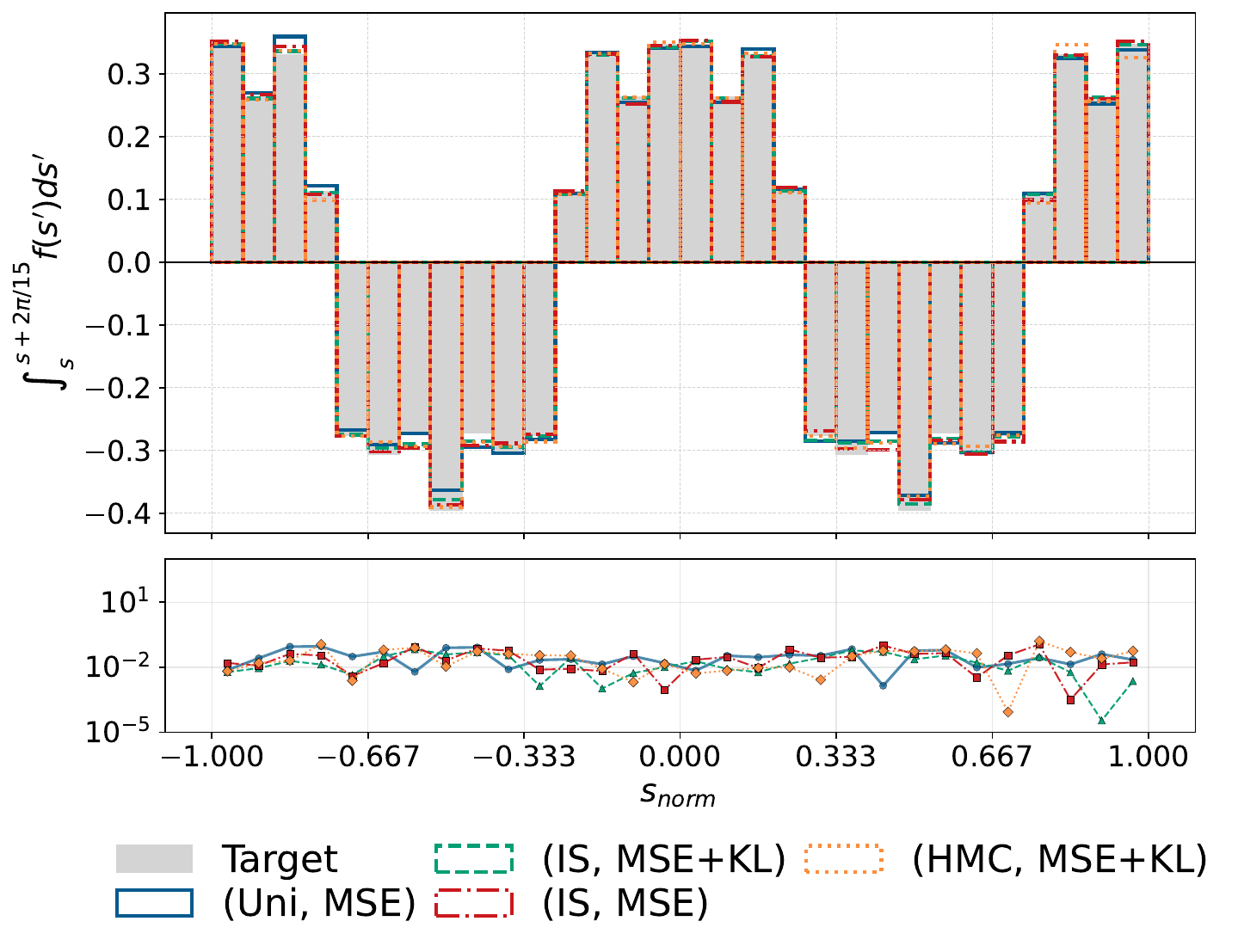}
    \par (b) Integral result in the subintervals
  \end{minipage}
  \begin{minipage}[t]{0.33\textwidth}
    \includegraphics[width=\linewidth,
                     height=0.28\textheight, keepaspectratio]
      {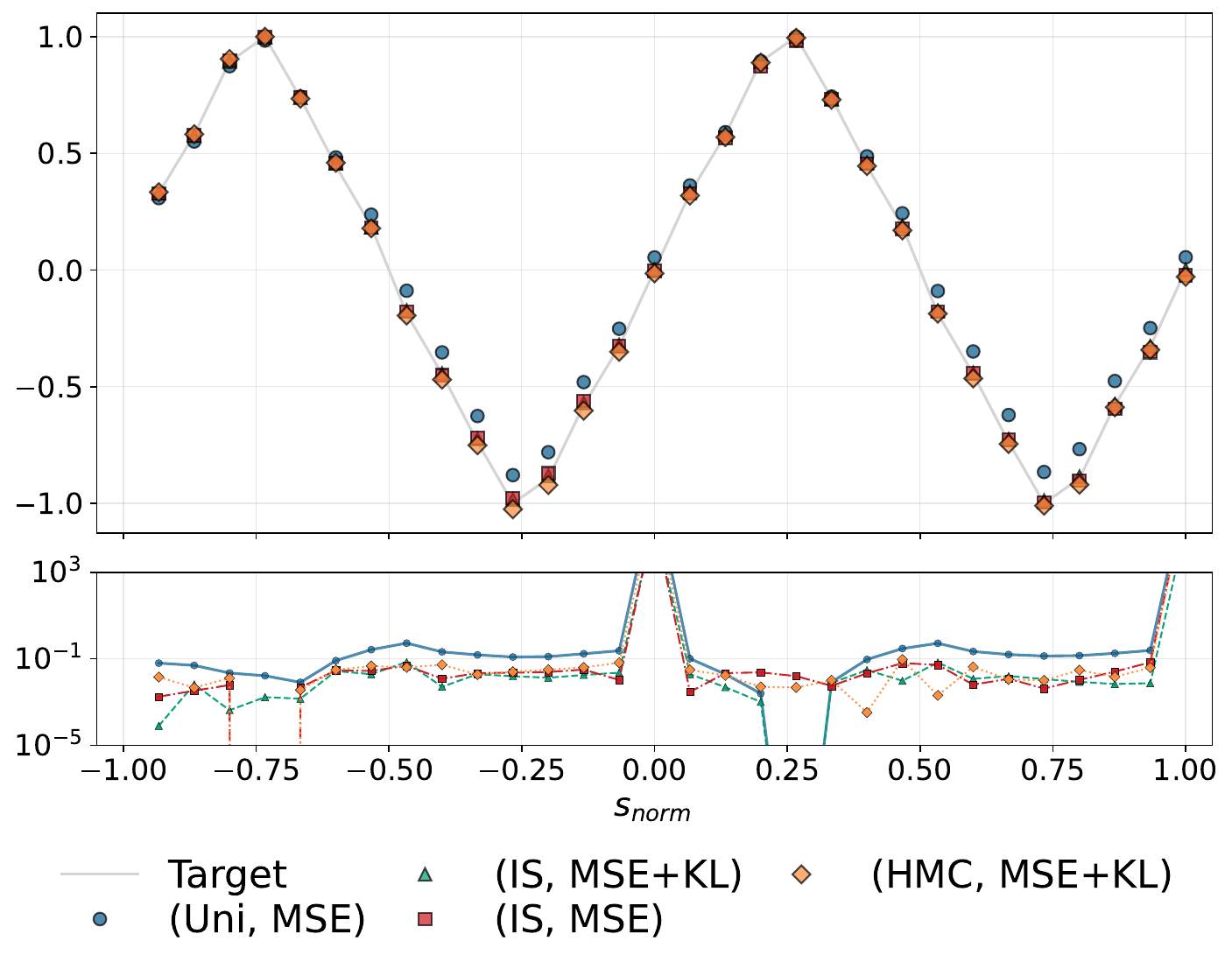}
    \par (c) Cumulative integral normalized result
  \end{minipage}
  \caption{
    QNN simulation results with the CPF example for the (Uniform, MSE) and three configurations with the lowest $W_1$ over the range is $s \in [-2\pi, 2\pi]$ with the normalized variable $s_{\text{norm}}$. 
    (a) derivative fit $q(x;\theta)$ versus \ target $f(x)$ (bottom strips: pointwise relative error, log scale); 
    (b) sub-interval integrals over $[0,\pi/2]$, $[-\pi/2,\pi/2]$, and $[-\pi,\pi]$; 
    (c) normalized cumulative integral. 
    The curves show the target, the baseline, and the three top configurations. 
    The numerical estimations ($R^2$, $W_1$, and interval errors) are summarized in Table~\ref{tab:CPF_result}.
    In panel (c), sharp peaks appear because the reference value is $0$ at $s = 0$ and $2\pi$. 
    At $s = \pi/2$, the (IS, MSE+KL) configuration exhibits particularly high accuracy in reproducing the cumulative integral relative to the reference.}
  \label{fig:CPF_result}
\end{figure*}

For these metrics, a higher $R^2$ value implies a more reliable result, and a lower $W_1$ distance points to a better model prediction along the training region.
We first evaluate the $R^2$ score for the models, and then compare the $W_1$ distance to determine the optimal data sampling method and loss function. 
Also, we investigate three subintervals containing the singular structure to clarify the integral performance around them.
Furthermore, the relative error for several points is also shown in the lower panels of Fig.~\ref{fig:CPF_result}-\ref{fig:bw_result} to see the error for each data point.
Since the QNN model achieves the best $R^2$ score, the discussion in this section centers on its results.
While the corresponding results of alternative models are presented in Appendix~\ref{appendix:result}.

\begin{table*}[t]
\setlength{\tabcolsep}{16pt}
\renewcommand{\arraystretch}{1.2}
\centering
\begin{tabular}{c|c|c|c|c|c}
\hline \hline
Training Strategy           & $R^2$ (Deriv.)     & $W_1$ (Int.) & $[0, \pi/2]$     & $[-\pi/2, \pi/2]$   & $[-\pi, \pi]$  \\ \hline\hline
(Uni., MSE)                  & 0.9972             & 0.0635       & 1.0695 (0.22\%)  & 2.1251 (0.42\%)     & 0.0009         \\ \hline
(IS, $\mathrm{MSE+KL}$)      & 0.9982             & 0.0062       & 1.0707 (0.33\%)  & 2.1293 (0.22\%)     & 0.0015         \\ \hline
(IS, MSE)                    & 0.9977             & 0.0100       & 1.0679 (0.07\%)  & 2.1256 (0.40\%)     & -0.0055        \\ \hline
(HMC, $\mathrm{MSE+KL}$ )    & 0.9975             & 0.0123       & 1.0684 (0.12\%)  & 2.1397 (0.25\%)     & 0.0005         \\ \hline\hline
\end{tabular}
\caption{
QNN simulation results for the CPF example ((Uniform, MSE) and the three best configurations depend on $W_1$). 
The first two columns report derivative fidelity $R^2$ and cumulative–integral alignment $W_1$. 
The last three columns give the predicted sub-interval integrals, where percentages in parentheses denote relative errors with respect to the analytic reference values ($1.0671$ for $[0, \pi/2]$ and $2.1342$ for $[-\pi/2, \pi/2]$). 
For the $[-\pi, \pi]$ interval, the predicted value is only reported because the reference value is 0.}
\label{tab:CPF_result}
\end{table*}

\subsection{Benchmark studies with exotic functions}

\subsubsection{Complex Periodic Function}\label{sec:CPF}

First, we evaluate our models on the Complex Periodic Function (CPF), defined as
\begin{equation}
    f_{\mathrm{CPF}}(s) = \cos\!\big(s + 0.5\,\sin(4s)\big),
\end{equation}
a phase modulated periodic signal, which combines a smooth, global baseline with rapid, local oscillations induced by nonlinear phase structure.
The CPF gives a non-uniform curvature and a spatially varying oscillation density across the domain.
We train the function over the interval $s \in [s_{\min}, s_{\max}]$, with $s_{\min} = -2\pi$ and $s_{\max} = 2\pi$, to assess its ability to capture the phase structure.
For convenience, we use the normalized variable $s_{\text{norm}} \equiv \frac{2s - (s_{\max} + s_{min})}{s_{\max} - s_{\min}}$, with $s_{\text{norm}} \in [-1, 1]$ and adopt it as the input variable throughout the training process.
The same normalization is applied to the benchmark examples presented in this study.

This function represents a simple toy example for the interference patterns in LHC scattering amplitudes, where distinct energy scales arise from superpositions of intermediate states or higher order perturbative effects.
Accurate numerical integration is required to resolve both the broad interference envelope and the localized, high-frequency oscillations that determine the final differential cross-section predictions.
Here, CPF provides a benchmark designed to assess a model’s ability to capture both global trends and localized high frequency features simultaneously.

Table~\ref{tab:CPF_result} shows that the three configurations with the lowest $W_1$ distances achieve high pointwise accuracy in approximating the derivative, with $R^2$ score above 0.9972.
Fig.~\ref{fig:CPF_result} (a) further shows that the models capture the local features of the integrand, closely following the target.
The $W_1$ distance and the errors on particular subintervals, however, show pronounced variations between the strategies.
Accurate integration is ultimately governed by the extent to which the model captures the global structure of the antiderivative, a process strongly influenced by the choice of sampling scheme and loss function.

With the lowest $W_1$ distance (0.0062) and the highest $R^2$ score (0.9982), (IS, MSE+KL) produces the most accurate results based on the performance metrics, and thus achieves the lowest error among all tested configurations as shown in Table~\ref{tab:CPF_result}. 
In addition, it shows better results for the subinterval integrals and the cumulative integral behavior as shown in Fig.~\ref{fig:CPF_result} (b) and (c)
The combination of (IS, MSE) with \(W_1=0.0100\) achieves the highest accuracy of the integral over $[0,\pi/2]$.
In contrast, the (HMC, MSE+KL), with \(W_1=0.0123\), provides a comparable level of accuracy and shows better results for the specific subintervals.

\begin{figure*}[t]
  \centering
  \begin{minipage}[t]{0.33\textwidth}
    \includegraphics[width=\linewidth, keepaspectratio]
      {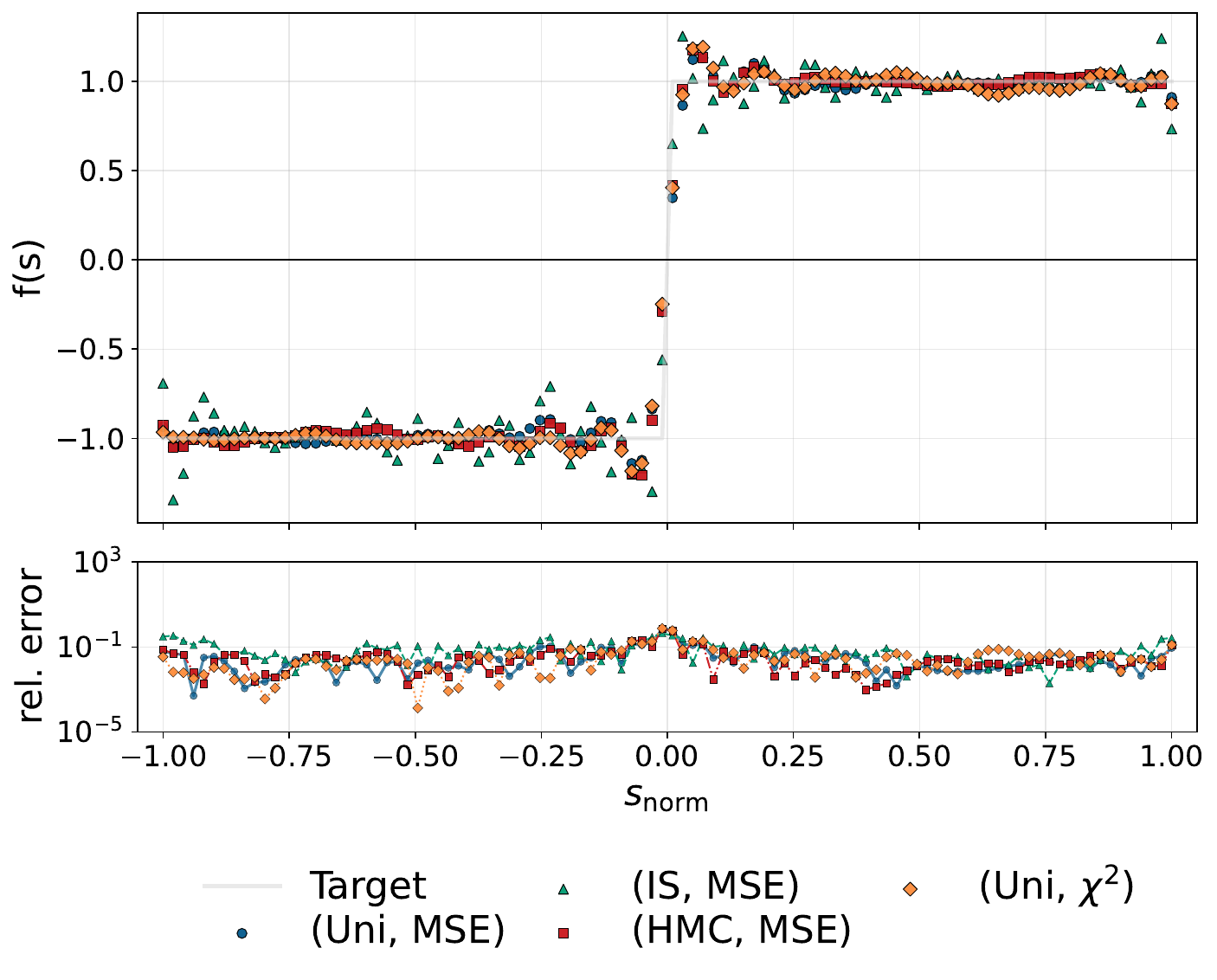}
    \par (a) Training result
  \end{minipage}
  \begin{minipage}[t]{0.33\textwidth}
    \includegraphics[width=\linewidth, keepaspectratio]
      {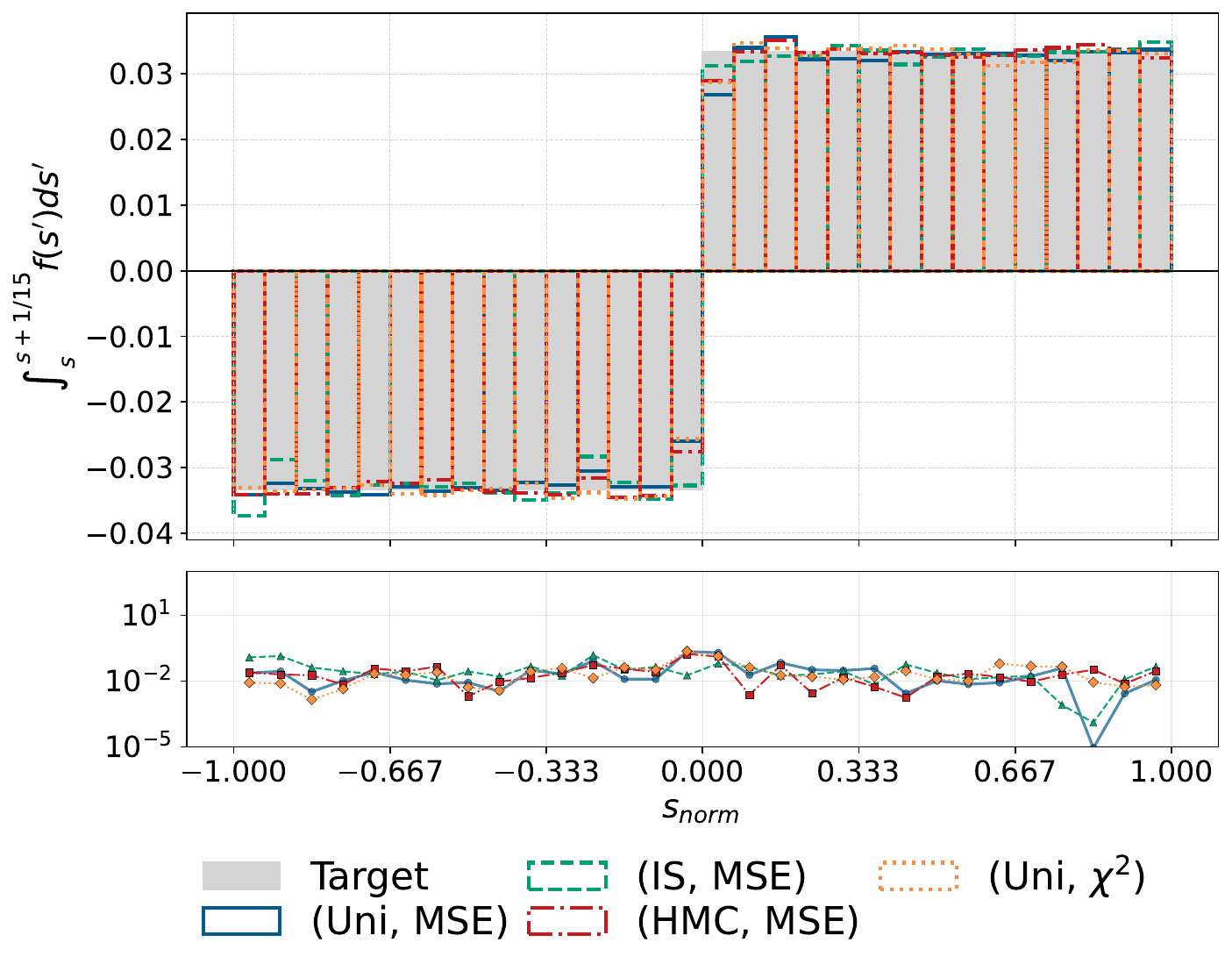}
    \par (b) Integral result in the subintervals
  \end{minipage}
  \begin{minipage}[t]{0.33\textwidth}
    \includegraphics[width=\linewidth, keepaspectratio]
      {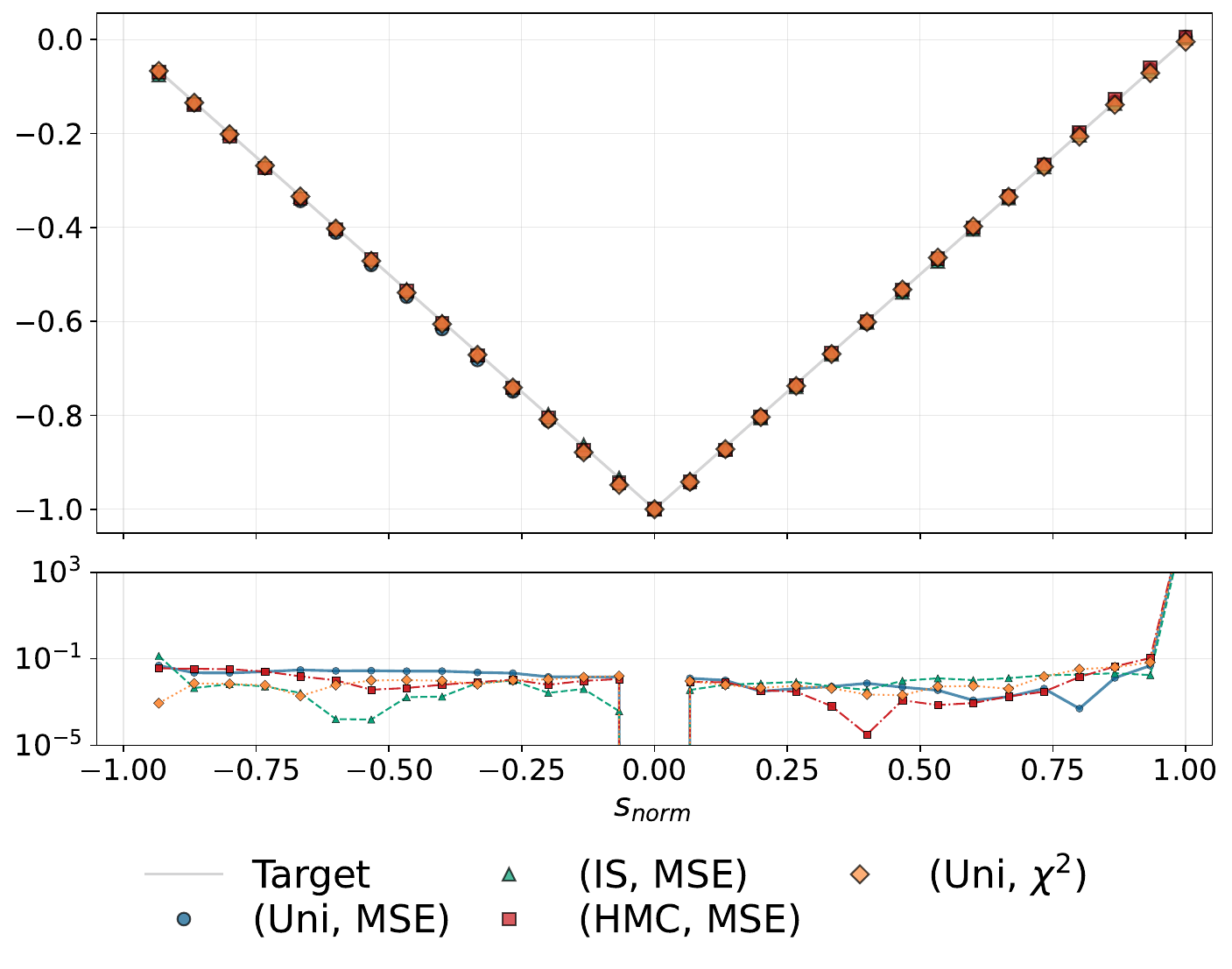}
    \par (c) Cumulative integral normalized result
  \end{minipage}
  \caption{QNN simulation results with the STEP example for the (Uniform,MSE) and three configurations with the lowest $W_1$ over the range is $s \equiv s_{\text{norm}} \in [-1, 1]$.
   (a) derivative fit $q(x;\theta)$ versus \ target $f(x)$ (bottom strips: pointwise relative error, log scale);
   (b) sub-interval integrals over $[-0.5,0]$, $[0,0.5]$, and $[-0.5,0.5]$;
   (c) normalized cumulative integral.
   Curves are composed with the same configurations as in Fig.~\ref{fig:CPF_result}.
   Numerical metrics ($R^2$, $W_1$, and interval errors) are summarized in Table~\ref{tab:step_result} (full QNN results in Table~\ref{tab:qnn_step_result}).
   In panel (c), a peak appears at $s = 1$ due to the vanishing reference value, while at $s = 0$ the prediction coincides with the reference as a result of the normalization procedure.
}
  \label{fig:step_result}
\end{figure*}

\begin{table*}[t]
\setlength{\tabcolsep}{15pt}
\renewcommand{\arraystretch}{1.2}
\centering
\begin{tabular}{c|c|c|c|c|c}
\hline \hline
Training Strategy           & $R^2$ (Deriv.)     & $W_1$ (Int.) & $[0, 0.5]$       & $[-0.5, 0]$         & $[-0.5, 0.5]$  \\ \hline\hline
(Uni., MSE)                  & 0.9882             & 0.0067       & 0.2429 (2.83\%)  & -0.2370 (5.19\%)    & 0.0058         \\ \hline
(IS, MSE)                    & 0.9856             & 0.0031       & 0.2443 (2.28\%)  & -0.2464 (1.40\%)    & -0.0021        \\ \hline
(HMC, MSE)                   & 0.9890             & 0.0040       & 0.2474 (1.03\%)  & -0.2460 (1.60\%)    & 0.0013         \\ \hline 
(Uni., $\chi^2$  Loss)       & 0.9879             & 0.0043       & 0.2492 (0.32\%)  & -0.2453 (1.88\%)    & 0.0039        \\ \hline\hline
\end{tabular}
\caption{
QNN simulation results for the STEP example ((Uniform, MSE) and the three best configurations ranked by $W_1$).
The first two columns report derivative fidelity ($R^2$) and cumulative–integral alignment ($W_1$).
The last three columns list the predicted sub-interval integrals, where percentages in parentheses denote relative errors with respect to the analytic reference values ($0.25$ for $[0,0.5]$ and $-0.25$ for $[-0.5,0]$).
For the final interval, $[-0.5, 0.5]$, the predicted value is only reported because its reference value is $0$.}
\label{tab:step_result}
\end{table*}

\subsubsection{Step Function}\label{sec:Step}

We test our network on the step function (STEP),
\begin{equation}
    f_{\mathrm{STEP}}(x)=
    \begin{cases}
        -0.5, & x<0,\\
        0.5, & x\ge 0,
    \end{cases}
\end{equation}
which has a discontinuity at \(x=0\). 
The step function serves as a benchmark for a model's robustness to non-smooth features, testing its ability to maintain integration accuracy across a domain containing a discontinuity.
We train the function over the interval $s \in [-1, 1]$, so that the data point $s$ is itself the normalized training variables $s_{\text{norm}}$.

The QNN performance on the step function is summarized in Fig.~\ref{fig:step_result} and Table~\ref{tab:step_result}, which compares the three top-performing configurations on $W_1$ distance to the (Uniform, MSE).
With $R^2$ scores above 0.9856, every tested configuration is able to learn the discontinuous behavior with high local fidelity.
However, a significant difference appears at the integral level.
With the lowest $W_1$ distance ($0.0031$) and the most competitive $R^2$ of $0.9856$, the (IS, MSE) combination exhibits the best global accuracy.
Interestingly, with a relatively low $W_1$ of $0.0040$, the (HMC, MSE) configuration obtained the highest derivative fidelity ($R^2 = 0.9890$).

The different training methods show complementary strengths, according to the subinterval error analysis.
Overall, the base approach (Uniform, MSE) stays constant, but it exhibits higher error close to the discontinuity, rising to 5.19\% on $[ -0.5, 0]$.
By contrast, (IS, MSE) shows a better approximation to the flat interval, reducing the error to 1.40\% over $[ -0.5, 0]$.
With the lowest error of 1.02\% on $[0,0.5]$, the (HMC, MSE) performs similarly on the positive subinterval.
Additionally, it performs well globally, obtaining one of the lowest absolute errors across the entire interval $[-0.5,0.5]$ ($|\Delta I|=0.0013$).
The (Uniform, $\chi^2$) reproduces the post-step level most accurately on [0,0.5], with an error of 0.32\%, but yields a slightly higher $W_1$.
Fig.~\ref{fig:step_result} (a) shows that all approaches exhibit very little overshoot at the discontinuity.
However, compared to  (Uniform, MSE) combination, the adaptive sampling strategies (IS and HMC) yield smoother residual distributions. 

\begin{figure*}[t]
  \centering
  \begin{minipage}[t]{0.33\textwidth}
    \includegraphics[width=\linewidth, height=0.28\textheight, keepaspectratio]
      {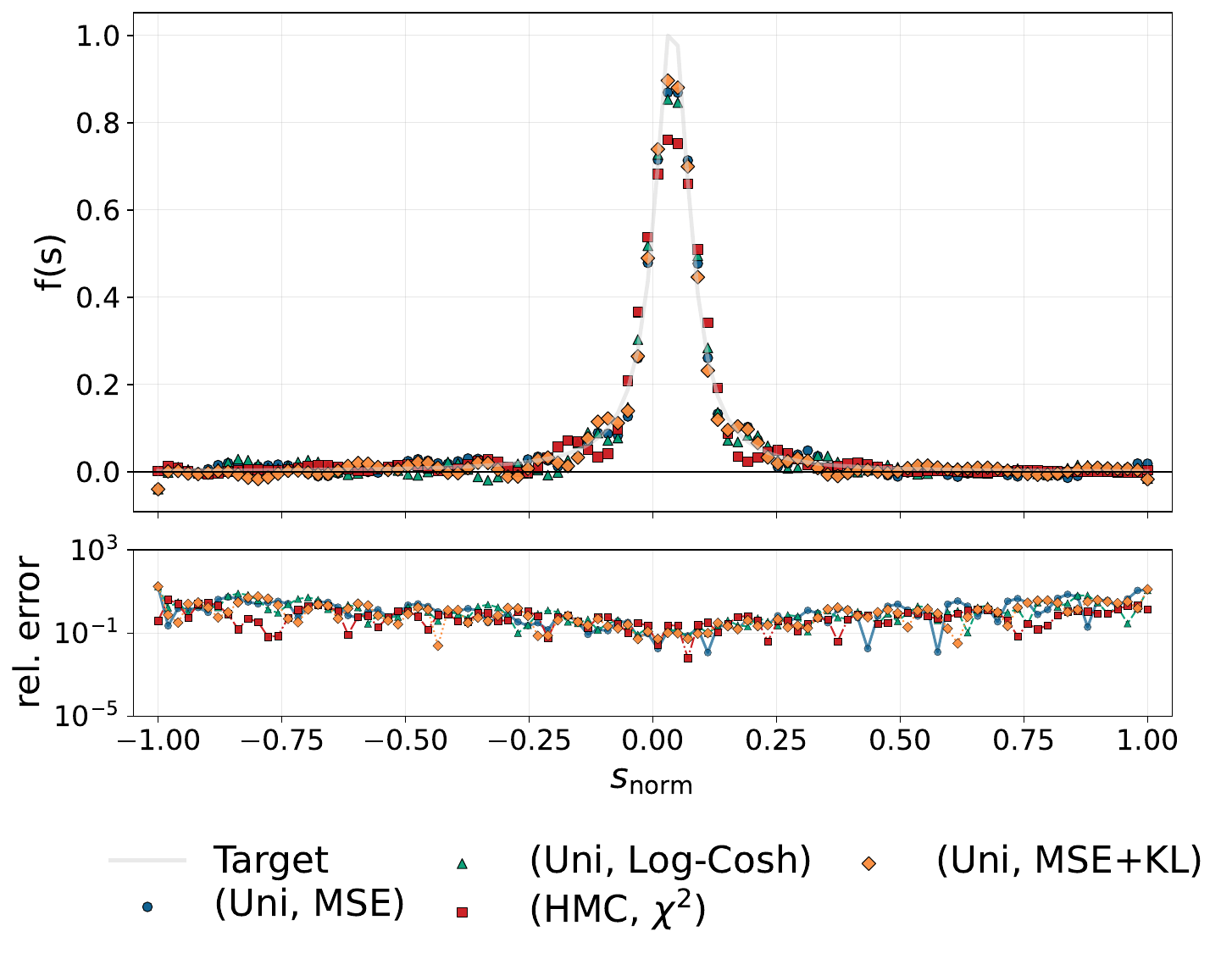}
    \par (a) Training result
  \end{minipage}
  \begin{minipage}[t]{0.33\textwidth}
    \includegraphics[width=\linewidth, height=0.28\textheight, keepaspectratio]
      {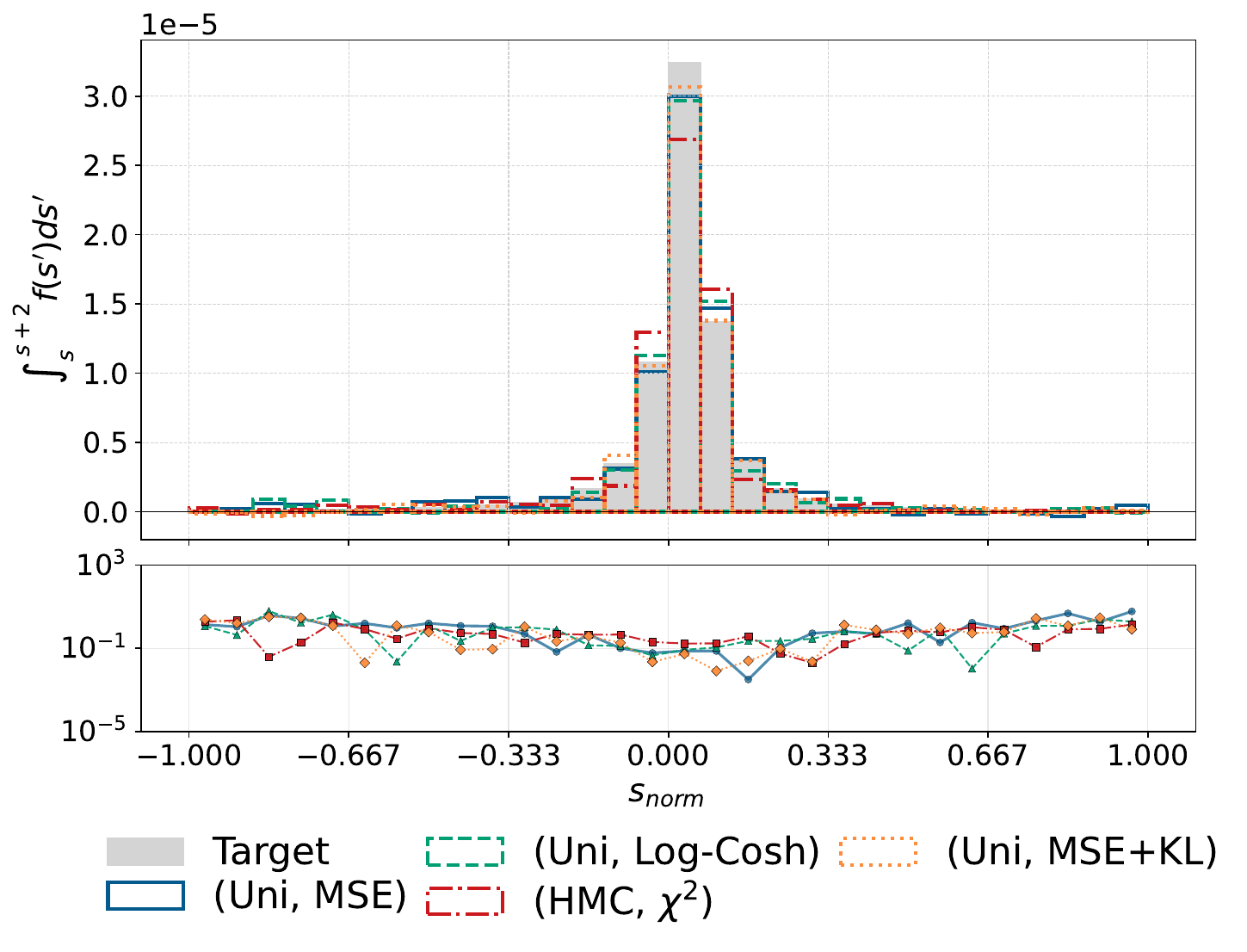}
    \par (b) Integral result in the subintervals
  \end{minipage}
  \begin{minipage}[t]{0.33\textwidth}
    \includegraphics[width=\linewidth, height=0.28\textheight, keepaspectratio]
      {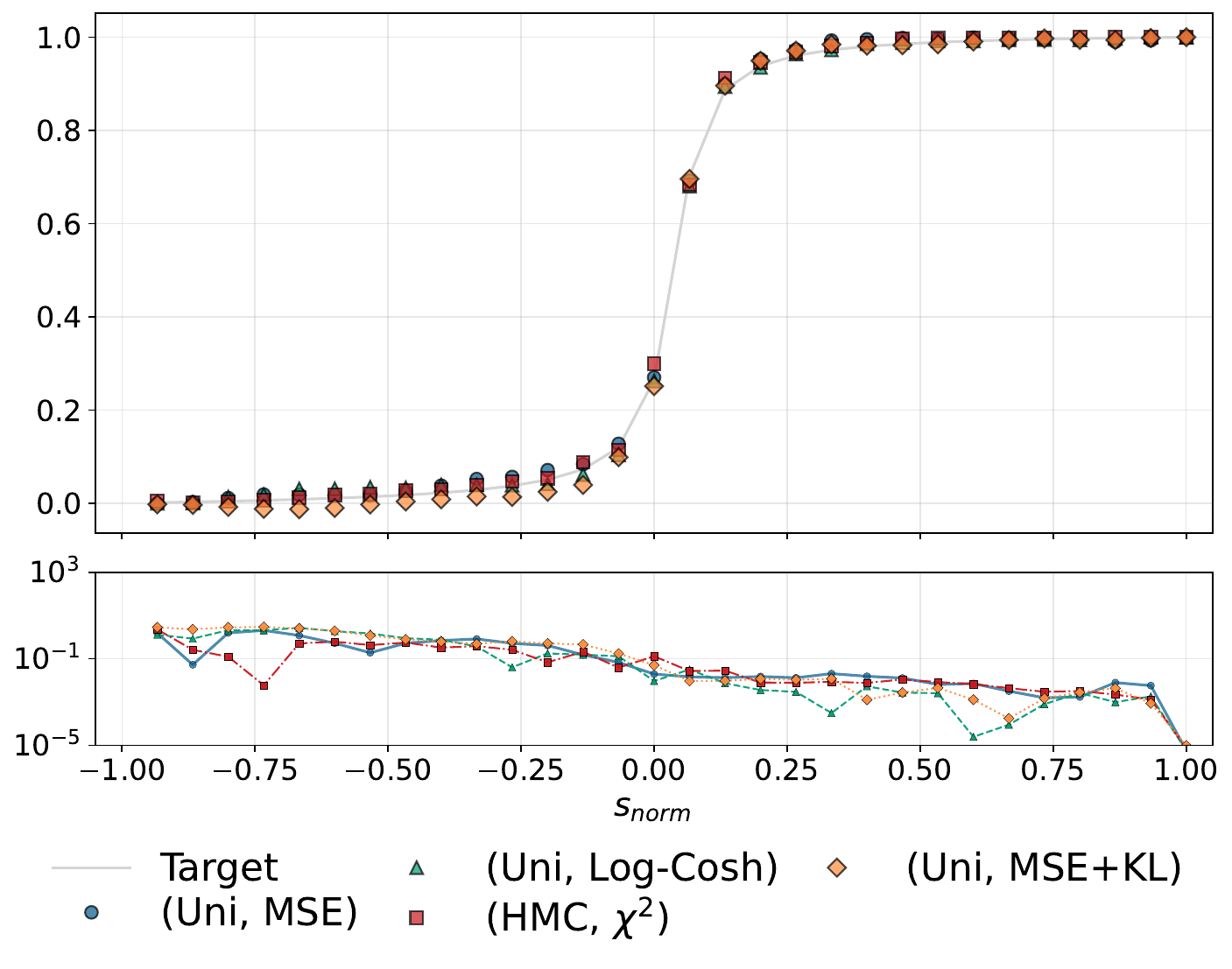}
    \par (c) Cumulative integral normalized result
  \end{minipage}
\caption{
QNN simulation results with the BW example for the (Uniform, MSE) and three configurations with the lowest $W_1$ over the range $s \in [60, 120]$ with the normalized variable $s_{\text{norm}} \in [-1, 1]$.
Panels: 
(a) derivative fit $q(s;\theta)$ vs.\ target $f(s)$ (bottom strips: pointwise relative error, log scale);
(b) sub-interval integrals over $[M_Z\pm3\Gamma]$, $[M_Z\pm5\Gamma]$, and $[M_Z\pm10\Gamma]$;
(c) normalized cumulative integral.
Curves is composed with the same configuration as before in Fig.~\ref{fig:CPF_result}.
Numerical metrics ($R^2$, $W_1$, and interval errors) are summarized in Table~\ref{tab:BW_result} (full QNN results in Table~\ref{tab:qnn_bw_result}).
}
\label{fig:bw_result}
\end{figure*}

\begin{table*}[t]
\setlength{\tabcolsep}{14pt}
\renewcommand{\arraystretch}{1.2}
\centering
\begin{tabular}{c|c|c|c|c|c}
\hline \hline
Training Strategy     & $R^2$ (Deriv.)     & $W_1$ (Int.) & $[M_Z \pm 3\Gamma]$ & $[M_Z \pm 5\Gamma]$ & $[M_Z\pm 10\Gamma]$  \\ \hline\hline
(Uni., MSE)            & 0.9804             & 0.0084       & 6.4476 (5.07\%)     & 6.7863 (4.57\%)     & 6.9471 (5.58\%)      \\ \hline
(Uni., Log-Cosh)       & 0.9732             & 0.0074       & 6.5534 (3.51\%)     & 6.7091 (5.65\%)     & 6.9818 (5.11\%)      \\ \hline
(HMC, $\chi^2$  Loss)  & 0.9472             & 0.0076       & 6.4752 (4.66\%)     & 6.7543 (5.02\%)     & 6.9806 (5.13\%)      \\ \hline 
(Uni., MSE+KL)         & 0.9848             & 0.0106       & 6.5748 (3.20\%)     & 6.7207 (5.49\%)     & 6.9180 (5.98\%)      \\ \hline\hline
\end{tabular}
\caption{
QNN simulation results for the BW example ((Uniform,MSE) and the three best configurations ranked by $W_1$).
The first two columns report derivative fidelity ($R^2$) and cumulative–integral alignment ($W_1$).
The last three columns list the predicted integrals over $[M_Z\pm3\Gamma]$, $[M_Z\pm5\Gamma]$, and $[M_Z\pm10\Gamma]$.
The percentages in parentheses denote relative errors with respect to the analytic Breit–Wigner integrals for each interval: $6.7924\times 10^{-5}$, $7.1113\times 10^{-5}$, and $7.3582\times 10^{-5}$.
All integral entries in the table are reported with a common scaling factor of $\times 10^{-5}$.
}
\label{tab:BW_result}
\end{table*}

\subsection{Breit-Wigner Distribution for Z boson}\label{sec:BW}

For the application for the high energy physics, we evaluate our models on the relativistic Breit-Wigner distribution (BW) for the $Z$ boson, defined as
\begin{equation}
    f_{\mathrm{BW}}(s) = \frac{1}{(s^2 - M_Z^2)^2 + M_Z^2 \Gamma^2},
\end{equation}
where $s$ is the center-of-mass energy, $M \simeq 91.1876$~GeV is the Z~boson mass, $\Gamma \simeq 2.4952$~GeV is its decay width.
For this distribution, the training is performed over the range $s \in [60, 120]$, using the normalized input $s_{\text{norm}}$ introduced before.
The BW distribution agrees with the typical resonance curve of an unstable particle in high energy physics, within the chosen training range.
An accurate integration of the BW distribution over a finite mass range is necessary in order to extract significant physical quantities, including precise resonance parameters and total cross-sections.
As a result, the localized resonance of Z boson is a particularly challenging case, since it needs to accurately reproduce the sharp peak intensity while maintaining numerical stability of the integration throughout the whole domain.

The three best combinations for the integral behavior are compared with the (Uniform, MSE) in Fig.~\ref{fig:bw_result}, and their performance metrics are summarized in Table~\ref{tab:BW_result}. 
The (Uniform, MSE) shows a reliable training result and integral behavior with $R^2=0.9804, W_1=0.0084$
The other three learning strategies also achieve a high $R^2$ score, which makes it valid to implement QuInt-Net for BW distribution.
Specifically, (Uniform, Log-Cosh) achieves the lowest $W_1$ distance (0.0074), which performs the best global integral stability compared to the target integral.
It shows a relative error below $6\%$ for the three subintervals, while using the regularized loss function serves a better result in the narrow interval, reducing the error below $3 \sim 4 \%$.

\begin{table*}[ht]
\setlength{\tabcolsep}{18pt}
\renewcommand{\arraystretch}{1.2}
\centering
\begin{tabular}{c|c|c|c|c}
\hline \hline 
Sampling              & Loss     & Gate Error                                 & Bit Flip                      & Depolarizing     \\ \hline\hline
                      & MSE      & $6.9125 (\pm 0.0755) (6.05\%)$             & $7.1493 (2.83\%)$             & $7.0412 (4.30\%)$    \\ \cline{2-5}
                      & Chisqr   & $5.8253 (\pm 0.0561) (20.83\%)$            & $5.7385 (22.01\%)$            & $5.7283 (22.15\%)$            \\ \cline{2-5}
                      & Log-Cosh & $7.1272 (\pm 0.0598) (3.13\%)$             & $7.0292 (4.47\%)$             & $7.0305 (4.45\%)$             \\ \cline{2-5}
\multirow{-4}{*}{Uni} & MSE+KL   & $\mathbf{7.2659 (\pm 0.0711) (1.25\%)}$    & $6.8596 (6.77\%)$             & $\mathbf{7.3366 (0.29\%)}$    \\ \hline
                      & MSE      & $6.7596 (\pm 0.760) (8.13\%)$              & $7.1136 (3.32\%)$             & $7.0138 (4.68\%)$             \\ \cline{2-5}
                      & Chisqr   & $6.6463 (\pm 0.0483) (9.67\%)$             & $6.8816 (6.47\%)$             & $6.9516 (5.52\%)$             \\ \cline{2-5}
                      & Log-Cosh & $6.6472 (\pm 0.0947) (9.66\%)$             & $7.2050 (2.08\%)$             & $6.7378 (8.43\%)$             \\ \cline{2-5}
\multirow{-4}{*}{IS}  & MSE+KL   & $6.6799 (\pm 0.1151) (9.21\%)$             & $6.9450 (5.61\%)$             & $6.9068 (6.13\%)$             \\ \hline
                      & MSE      & $6.9905 (\pm 0.0817) (4.99\%)$             & $6.9769 (5.18\%)$             & $7.0133 (4.68\%)$             \\ \cline{2-5}
                      & Chisqr   & $6.8548 (\pm 0.0507) (6.84\%)$             & $6.8018 (7.56\%)$             & $6.7859 (7.77\%)$             \\ \cline{2-5}
                      & Log-Cosh & $6.8899 (\pm 0.0933) (6.36\%)$             & $\mathbf{7.2212 (1.86\%)}$    & $6.7862 (7.77\%)$             \\ \cline{2-5}
\multirow{-4}{*}{HMC} & MSE+KL   & $7.5085 (\pm 0.1215) (2.04\%)$             & $6.7414 (8.38\%)$             & $7.0583 (4.07\%)$ \\ \hline\hline
\end{tabular}
\caption{
QNN simulation results for the optimized circuits, taking the BW in the range of $[M_Z \pm 10\Gamma]$ as an example, under noise with $\delta = p = 0.1\%$. The predicted integral values are reported in units of $10^{-5}$ averaged over 1,000 runs and their standard deviations, along with the corresponding error rates ($\pm$ their standard deviations) for three different sources of noise. The percentages in parentheses are relative errors with respect to the reference value $7.3582 \times 10^{-5}$.
The values are emphasized with \textbf{bold} for the best prediction.
}
\label{tab:qnn_bw_noise}
\end{table*}

We also evaluated the Breit-Wigner integral over the interval $[M_Z \pm 10\Gamma]$ in the presence of three representative noise models: gate error, bit-flip error, and depolarizing noise, in order to examine the performance of our QNN model under hardware noise. The results are summarized in Table~\ref{tab:qnn_bw_noise}.
Gate error is implemented as a parameter deviation of the form $\theta + \delta \phi$, with $\delta = 0.1\%$ and $\phi \sim \mathcal{N}(0,1)$. 
In accordance with single-qubit error rates documented on superconducting and trapped-ion platforms~\cite{arute2019quantum, wright2019benchmarking}, we used a $0.1\%$ probability of error occurrence for bit-flip and depolarizing channels. 

The noise levels offer a useful test of the VQC's stability to hardware flaws since they correspond to the low-error regimes expected for near-term quantum devices~\cite{arute2019quantum, wright2019benchmarking}.
We operate the QuInt-Net framework in a noise-aware way, incorporating the same noise models used in the final evaluation during the training loop's forward pass.
In particular, density matrix operations are used to simulate the evolution of the circuit for bit-flip and depolarizing channels.
In order to prevent overfitting to a perfect, noiseless simulator and to increase the model's resilience to hardware-level disturbances, the loss function is then minimized over an average of the stochastic realizations.

Under the gate error, the (Uniform, MSE+KL) makes the best prediction with just $1.25\%$, whereas the (HMC, MSE+KL) consistently predicts $2.04\%$ .
For bit-flip, the (HMC, Log-Cosh) and (IS, Log-Cosh) performs the lowest errors ($1.86\%$ and $2.08\%$, respectively). 
The (Uniform, MSE+KL) shows an exceptionally low relative error of $0.29\%$ in depolarizing channel, outperforming other learning combinations with errors ranging from $4\% \sim 20\%$.

As shown in Table~\ref{tab:qnn_bw_noise}, the results confirm that although the QNN architecture is naturally robust to noise, the choice of loss function has an impact on the final integral calculation.
Our findings show that the MSE+KL and Log-Cosh losses offer protection against hardware errors.
In particular, the (HMC, Log-Cosh) approach works best against bit-flip noise, while the (Uniform, MSE+KL) configuration is most resilient to gate and depolarizing errors.
Taken together, these results suggest that pairing an adaptive sampling strategy with a regularized, noise-tolerant loss function is a promising approach for improving the robustness of QML applications on near-term devices.

\section{Conclusion and Discussion}\label{sec:Conclusion}

In this work, we introduce QuInt-Net, a variational circuit framework for numerical integration that approximates antiderivatives directly, enabling endpoint evaluation of definite integrals.
We use three benchmark cases to test the performance: (i) a highly oscillatory function, where conventional Monte Carlo methods suffer from high variance; (ii) a function with a discontinuity, to probe stability for a noncontinuous function; and (iii) the Breit–Wigner form, which features a singular resonance.

Because QuInt-Net is based on quantum machine learning, we investigate different strategies for preparing training data, including uniform sampling, importance sampling (IS), and Hamiltonian Monte Carlo (HMC). 
We also compared several loss functions ($\chi^2$-weighted, Log-Cosh, and MSE with KL regularization). 
The model output is evaluated by $R^2$ score and Wasserstein distance for the model approximation and prediction respectively. 
We found that IS combined with MSE+KL yields the lowest $W_1$ for oscillatory functions, while regulated loss strategies are advantageous for resonance-dominated function.
However, the overall gain remains limited.
The strategies, while beneficial in certain regimes, provide only incremental improvements and do not deliver a substantial enhancement in performance.

To further assess practical applicability, we investigate the robustness of the network under noise channels, including gate, bit-flip, and depolarizing at the $0.1\%$ level. 
For the Breit–Wigner case, the results show that the QNN model retains high fidelity, with the accuracy of integral estimates depending primarily on the choice of loss function. 
Among them, MSE+KL and Log-Cosh losses provide the most stable performance across noise sources.

Considering the different circuit designs, QNN offers the best overall performance over our considered benchmarks, QSP circuits improve the treatment of discontinuities by mitigating overshoot effects, and DQC1 circuits yield stable results for smooth integrands, owing to the long coherence time ($T_2$) of NMR platforms~\cite{lu2016nmr}, but their performance remains inferior to that of QNN realizations~\cite{kim2025expressivity}.

Overall, these results demonstrate the feasibility of variational quantum integration with diverse learning strategies. 
Extending the method to higher-dimensional integrals and implementing it on real hardware remains challenging and defines important directions for future work. 
Further studies will be needed to assess performance under realistic noise conditions, optimize resource requirements, and explore the applicability of QuInt-Net to phase-space integrals in collider physics.

\section{Acknowledgements}

This work is supported by the 2024 IonQ Spring Research Mentorship Program.
We thank Evgeny Epifanovsky for his valuable feedback.
KB is supported by the KIAS Individual Grant No. PG097601. KK is supported in part by the US DOE under Award No DE-SC0024407. MP is supported by the National
Research Foundation of Korea (NRF) grant funded by the Korea government (MSIT) (No.RS-2025-00564488)

\clearpage

\appendix

\section{Quantum Noise Environment : Mathematical Derivation}\label{appendix:quantum_noise}

Quantum circuits are inherently prone to errors due to imperfections in gate implementation and interactions with the environment. 
Here, we provide a brief overview of three fundamental noise models that are widely used to describe such errors in quantum computing: gate errors, bit flip errors, and depolarizing errors.

Gate errors arise when the physical implementation of quantum gates deviates from its ideal operation.
These operations are affected by calibration inaccuracies, pulse distortions, or fluctuations in control parameters, which cause gate errors and accumulate in deep circuits.
For example, unintended angle rotation for a single-qubit gate $R_\sigma(\theta)$ operates with a discrepancy like $R_\sigma(\theta + \delta\phi)$.
Therefore, the quantum algorithms with a large number of gate operations are hard to implement

Another common noise channel is the bit flip error, which describes the possibility of a qubit flipping its state from $\ket{0}$ to $\ket{1}$ or vice versa. 
This error can be modeled by the following Kraus operators:
\begin{equation}
    E_0 = \sqrt{1 - p} I, \quad E_1 = \sqrt{p} X,
\end{equation}
where $p$ denotes the probability of a bit flip occurring, $I$ is the identity operator, and $X$ is the Pauli-X matrix. 
Bit flip errors can physically arise from spurious electromagnetic fluctuations or unintended interactions between qubits and the environment.

The depolarizing channel represents a more uniform form of noise, where a qubit’s state loses its coherence and becomes closer to a maximally mixed state. 
This noise process can be mathematically described as:
\begin{equation}
    \mathcal{E}(\rho) = \frac{p}{2} I + (1 - p)\rho.
\end{equation}
The process can also be described using Kraus operators as:
\begin{equation}
    E_0 = \sqrt{1 - p} I, \quad E_i = \sqrt{\frac{p}{3}} P_i,
\end{equation}
where $P_1 = X$, $P_2 = Y$, and $P_3 = Z$ are the Pauli matrices. 
Depolarizing errors effectively contract the Bloch sphere uniformly, modeling scenarios where the qubit state partially or completely randomizes due to isotropic noise.

Understanding these noise models is crucial for analyzing the performance of quantum algorithms on NISQ devices~\cite{preskill2018quantum, horowitz2019quantum}.
While the models presented here are simplified, they provide essential insights into how errors manifest in real hardware and how error mitigation or correction techniques can be designed to compensate for these effects.

\section{Circuit Models and Detailed Results}\label{appendix:result}

\begin{figure}[t]
    \centering
    \begin{minipage}{\linewidth}
        \includegraphics[width=0.98\linewidth]{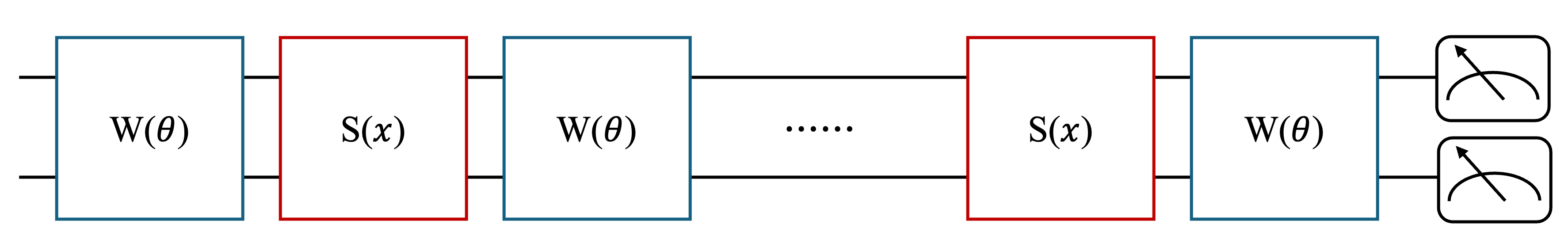}
        \par (a) Basic QNN Model.
    \end{minipage}
    \vspace{1em}
    
    \begin{minipage}{\linewidth}
        \centering
        \includegraphics[width=0.98\linewidth]{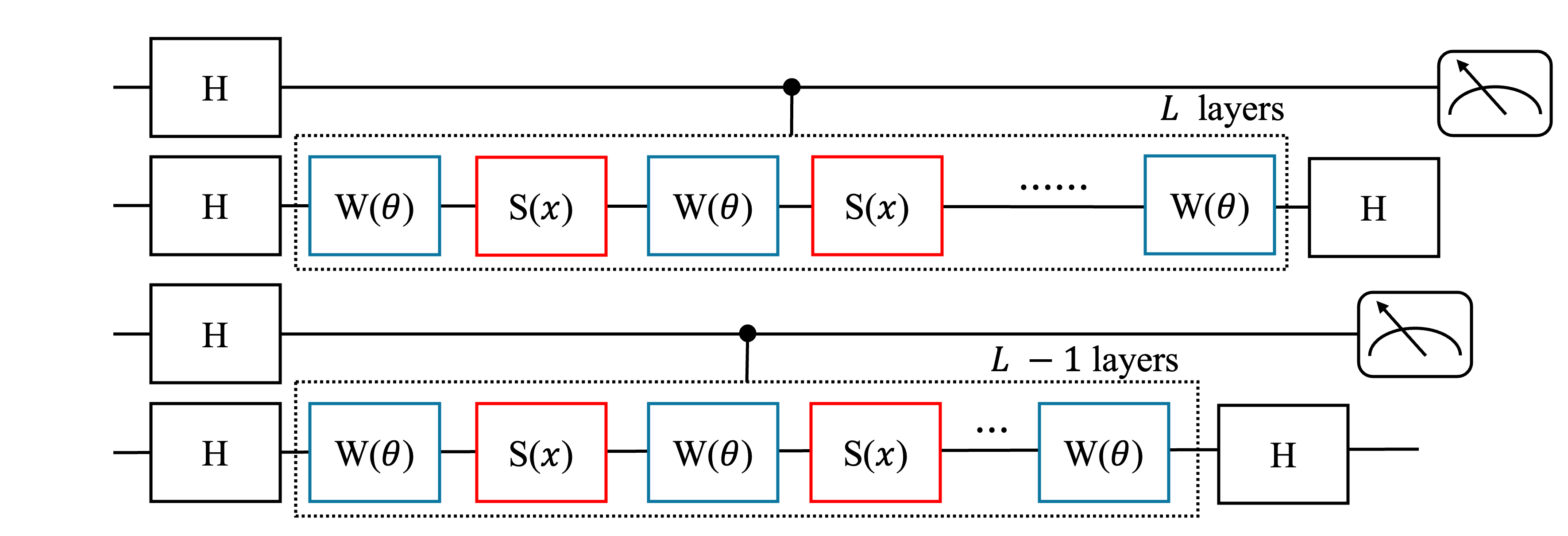}
        \par (b) QSP Model.
    \end{minipage}
    \vspace{1em}
    
    \begin{minipage}{\linewidth}
        \centering
        \includegraphics[width=0.98\linewidth]{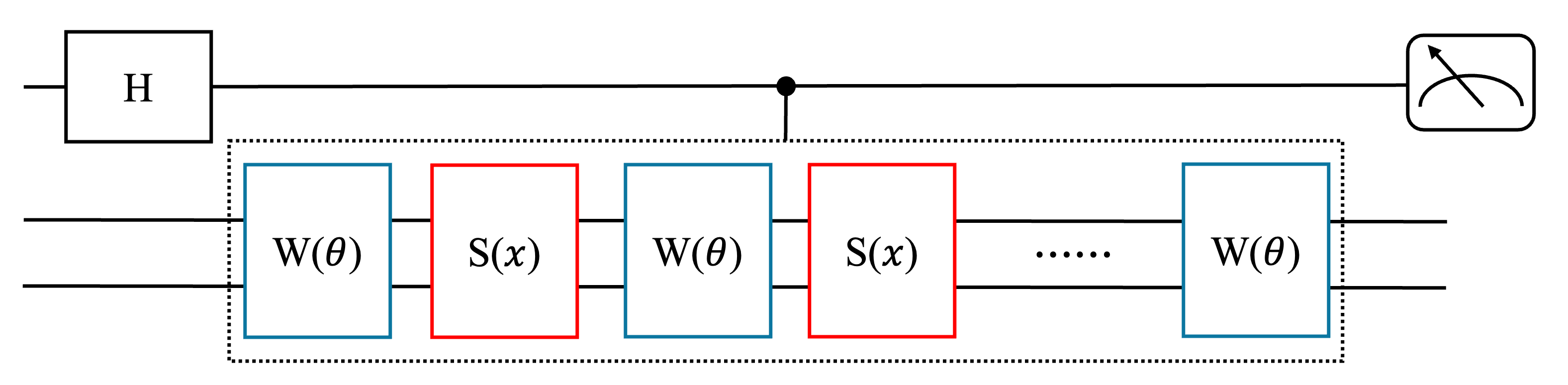}
        \par (c) DQC1 Model. 
    \end{minipage}
    \caption{Quantum circuit architectures used for QuInt-Net in this study. (a) The QNN model repeatedly embeds classical input data across circuit layers to approximate functions in a Fourier-like manner. (b) The QSP model used structured signal processing and phase rotation layers to implement polynomial transformations, measured via the Hadamard test. (c) The DQC1 model estimates the normalized trace of a unitary operator using a single clean qubit and a mixed-state register, enabling frequency-domain approximation.}
    \label{fig:Circuit}
\end{figure}

\begin{table}[th]
\setlength{\tabcolsep}{20pt}
\renewcommand{\arraystretch}{1.2}
\centering
\begin{tabular}{c|c|c}
\hline\hline
Model             & $P$                   & $E$               \\ \hline \hline
QNN               & $6L + 2$              & $2L$              \\ \hline
QSP               & $2L + 1$              & $2L - 1$          \\ \hline
DQC1              & $8L + 6$              & $2L$              \\ \hline\hline
\end{tabular}
\caption{
    The number of trainable parameters ($P$) and data encodings ($E$) for each circuit architecture are shown as a function of the number of embedding blocks $L$ (highlighted as red blocks in Fig.~\ref{fig:Circuit}). 
    While the DQC1 model has a larger parameters, this is due to its ability to support deeper circuits with minimal coherence requirements.}
\label{tab:compar_circuit}
\end{table}

We compare all three architectures and present comprehensive results for the alternative circuit models, QSP and DQC1, in addition to the primary QNN analysis. 
The number of data embedding repetitions is set to $L=10$ for consistency across models, separating the impact of input encoding from that of circuit design. 
The three architectures are depicted in Fig.~\ref{fig:Circuit}, and their trainable parameter counts and data encodings are compiled in Table~\ref{tab:compar_circuit}. 
A Hadamard test for QSP, the Pauli-$X$ expectation on the clean qubit for DQC1, and joint Pauli-$Z$ correlations for QNN are the output measurement schemes selected to correspond with each architecture. 
The ensuing subsections provide more thorough explanations and numerical findings.

\subsection{QNN Model}\label{appendix:qnn_result}

Using the data re-uploading strategy, which repeatedly embeds classical inputs across circuit layers, QNN serves as the main ansatz of QuInt-Net~\cite{perez2020data}. 
With the number of re-uploads corresponding to the effective Fourier modes represented, this mechanism functions as a quantum analogue of Fourier series expansion~\cite{schuld2021effect}.
The circuit is constructed as shown in \ref{fig:Circuit}(a) and the gate are designed like:
\begin{equation} 
\mathcal{U}_{\text{Fourier}}(x | \theta) = R_X(\theta_0) R_Z(\theta_1 \cdot x) R_Y(\theta_2) \, , 
\end{equation}
followed by entangling CZ gates to improve expressivity. 

We now expand upon the summaries presented in Section~\ref{sec:Result} by providing the full numerical results produced by the QNN model.
All combinations of sampling strategies (Uniform, IS, HMC) and loss functions (MSE, $\chi^2$, Log-Cosh, and MSE+KL) are included.
For the three benchmark functions—CPF, STEP, BW—we provide the derivative fidelity ($R^2$), global cumulative behavior ($W_1$), and sub-interval integrals compared with analytic references. 
The parentheses gives the relative error rates in Table~\ref{tab:qnn_cpf_result}--\ref{tab:qnn_bw_result}.

\begin{table*}[!h]
\setlength{\tabcolsep}{13pt}
\renewcommand{\arraystretch}{1.2}
\centering
\begin{tabular}{l|l|c|c|c|c|c}
\hline \hline 
Sampling              & Loss     & $R^2$              & $W_1$              & $[0, \pi/2]$                & $[\pi/2, \pi/2]$             & $[-\pi, \pi]$      \\ \hline\hline
                      & MSE      & 0.9972             & 0.0635             & 1.0695 (0.22\%)             & 2.1251 (0.42\%)              & \underline{0.0009} \\ \cline{2-7}
                      & Chisqr   & \underline{0.9980} & 0.0337             & 1.0810 (1.30\%)             & 2.1405 (0.29\%)              & 0.0284             \\ \cline{2-7}
                      & Log-Cosh & \textit{0.9977}    & 0.0433             & 1.0629 (0.39\%)             & 2.1601 (1.21\%)              & 0.0226             \\ \cline{2-7}
\multirow{-4}{*}{Uni} & MSE+KL   & 0.9975             & 0.0128             & 1.0648 (0.21\%)             & \textbf{2.1337 (0.02\%)}     & 0.0256             \\ \hline
                      & MSE      & \textit{0.9977}    & \underline{0.0100} & \textbf{1.0679 (0.07\%)}    & 2.1256 (0.40\%)              & -0.0055            \\ \cline{2-7}
                      & Chisqr   & 0.9973             & 0.0425             & 1.0698 (0.25\%)             & 2.1404 (0.28\%)              & 0.0555             \\ \cline{2-7}
                      & Log-Cosh & 0.9972             & 0.0314             & 1.0931 (2.43\%)             & 2.1782 (2.05\%)              & -0.0028            \\ \cline{2-7}
\multirow{-4}{*}{IS}  & MSE+KL   & \textbf{0.9982}    & \textbf{0.0062}    & 1.0707 (0.33\%)             & \textit{2.1293 (0.22\%)}     & \textit{0.0015}    \\ \hline
                      & MSE      & 0.9976             & 0.0208             & 1.0631 (0.37\%)             & 2.1284 (0.27\%)              & -0.0183            \\ \cline{2-7}
                      & Chisqr   & \textit{0.9977}    & 0.0144             & 1.0809 (1.29\%)             & 2.1519 (0.83\%)              & -0.0021            \\ \cline{2-7}
                      & Log-Cosh & 0.9970             & 0.0271             & \underline{1.0659 (0.10\%)} & \underline{2.1328 (0.06\%)}  & 0.0169             \\ \cline{2-7}
\multirow{-4}{*}{HMC} & MSE+KL   & 0.9975             & \textit{0.0123}    & \textit{1.0684 (0.11\%)}    & 2.1397 (0.25\%)              & \textbf{0.0005}    \\ \hline\hline
\end{tabular}
\caption{Complete QNN results for the CPF benchmark. 
Each entry reports derivative fidelity ($R^2$), cumulative-integral alignment ($W_1$), and predicted integrals over sub-intervals $[0,\pi/2]$, $[-\pi/2,\pi/2]$, and $[-\pi,\pi]$. 
The last three columns give the predicted sub-interval integrals, where percentages in parentheses denote relative errors with respect to the analytic reference values ($1.0671$ for $[0, \pi/2]$ and $2.1342$ for $[-\pi/2, \pi/2]$). 
For the $[-\pi, \pi]$ interval, the predicted value is only reported because the reference value is 0.
Here, the values are emphasized with \textbf{bold} for the best, \underline{underline} for the second and \textit{italic} for the third value.}
\label{tab:qnn_cpf_result}
\end{table*}

For the CPF example, Table~\ref{tab:qnn_cpf_result} presents the complete results of learning strategies (sampling and loss).
The QNN model achieves high learning result with $R^2 > 0.997$ confirming that the derivative of its output reliably capture the multi-scale oscillation structure.
However, the $W_1$ distance is distinguished across the combination.
(IS, MSE+KL) presents the best global integral behavior with $W_1 = 0.062$
In the subinterval integral results, (IS, MSE) has the lowest error $0.07\%$ in $[0, \pi/2]$,  while (Uniform, MSE+KL) has the lowest error $0.02\%$ in $[-\pi/2, pi/2]$.
Furthermore, (HMC, MSE+KL) has the lowest error $|\Delta I| = 0.0005$ in $[-\pi, \pi]$.
The (IS, Log-Cosh) shows robust performance by generating low errors in both intervals ($0.10\%$ and $0.06\%$) and (IS, MSE+KL) shows the third best result in $[-\pi/2, pi/2]$ and $[-\pi, \pi]$.
For the multi-oscillatory function, the result suggests that (IS, MSE+KL) could be more optimal.
The findings of the study indicate that it may be optimal to use the (IS, MSE+KL) for complex periodic functions for QuInt-Net.

\begin{table*}[!h]
\setlength{\tabcolsep}{13pt}
\renewcommand{\arraystretch}{1.2}
\centering
\begin{tabular}{l|l|c|c|c|c|c}
\hline \hline 
Sampling              & Loss     & $R^2$              & $W_1$              & $[0, 0.5]$                  & $[-0.5, 0]$                  & $[-0.5, 0.5]$       \\ \hline\hline
                      & MSE      & \underline{0.9882} & 0.0067             & 0.2429 (2.83\%)             & -0.2370 (5.19\%)             & 0.0058              \\ \cline{2-7}
                      & Chisqr   & \textit{0.9879}    & \textit{0.0043}    & \textbf{0.2492 (0.28\%)}    & \textit{-0.2453 (1.84\%)}    & 0.0039              \\ \cline{2-7}
                      & Log-Cosh & 0.9875             & 0.0089             & 0.2424 (3.02\%)             & -0.2410 (3.56\%)             & \textbf{0.0013}     \\ \cline{2-7}
\multirow{-4}{*}{Uni} & MSE+KL   & 0.9871             & 0.0098             & 0.2444 (2.21\%)             & -0.2396 (4.12\%)             & 0.0047              \\ \hline
                      & MSE      & 0.9856             & \textbf{0.0031}    & 0.2443 (2.27\%)             & \textbf{-0.2464 (1.40\%)}    & \underline{-0.0021} \\ \cline{2-7}
                      & Chisqr   & 0.9820             & 0.0057             & \textit{0.2523 (0.94\%)}    & -0.2453 (1.85\%)             & 0.0070              \\ \cline{2-7}
                      & Log-Cosh & 0.9862             & 0.0044             & 0.2451 (1.92\%)             & -0.2425 (2.97\%)             & \textit{0.0026}     \\ \cline{2-7}
\multirow{-4}{*}{IS}  & MSE+KL   & 0.9871             & 0.0080             & 0.2382 (4.70\%)             & -0.2444 (2.22\%)             & -0.0062             \\ \hline
                      & MSE      & \textbf{0.9890}    & \underline{0.0040} & 0.2474 (1.02\%)             & \underline{-0.2460 (1.57\%)} & \textbf{0.0013}     \\ \cline{2-7}
                      & Chisqr   & 0.9873             & 0.0078             & 0.2467 (1.29\%)             & -0.2401 (3.95\%)             & 0.0066              \\ \cline{2-7}
                      & Log-Cosh & 0.9874             & 0.0060             & \underline{0.2479 (0.80\%)} & -0.2430 (2.76\%)             & 0.0048              \\ \cline{2-7}
\multirow{-4}{*}{HMC} & MSE+KL   & 0.9830             & 0.0068             & 0.2446 (2.14\%)             & -0.2402 (3.90\%)             & 0.0043              \\ \hline\hline
\end{tabular}
\caption{Complete QNN results for the STEP function. 
Columns list $R^2$, $W_1$, and sub-interval integrals $[0,0.5]$, $[-0.5,0]$, and $[-0.5,0.5]$. 
The last three columns list the predicted sub-interval integrals, where percentages in parentheses denote relative errors with respect to the analytic reference values ($0.25$ for $[0,0.5]$ and $-0.25$ for $[-0.5,0]$).
For the final interval, $[-0.5, 0.5]$, the predicted value is only reported because its reference value is $0.$
The values are emphasized with \textbf{bold} for the best, \underline{underline} for the second, and \textit{italic} for the third value.}
\label{tab:qnn_step_result}
\end{table*}

Table~\ref{tab:qnn_step_result} presents the detailed results corresponding to the summary provided in Table~\ref{tab:step_result}.
It also presents high learning results for reliability with $R^2 > 0.982$, especially (HMC, MSE) shows the highest $R^2$ score with $0.9890$.
For the $W_1$ distance,  (IS, MSE) presents the lowest value, which means that it has good global integral behavior.
The (Uniform, $\chi^2$) has the lowest error $0.28\%$ in $[0,0.5]$ and (IS, MSE) achieves the best result in $[-0.5,0]$ with error $1.40\%$. 
For the subinterval $[-0.5, 0.5]$ that includes the discontinuous point, (Uniform, Log-Cosh) and (HMC, MSE) have the lowest error $|\Delta I| = 0.0013$.
The results suggest that (HMC, MSE) provides balanced results with the global integral performance and the integral including the singular structure.

\begin{table*}[!h]
\setlength{\tabcolsep}{13pt}
\renewcommand{\arraystretch}{1.2}
\centering
\begin{tabular}{l|l|c|c|c|c|c}
\hline \hline 
Sampling              & Loss     & $R^2$              & $W_1$              & $[M_Z \pm 3\Gamma]$         & $[M_Z \pm 5\Gamma]$         & $[M_Z \pm 10\Gamma]$        \\ \hline\hline
                      & MSE      & \textit{0.9804}    & \textit{0.0084}    & 6.4476 (5.07\%)             & 6.7863 (4.56\%)             & 6.9471 (5.58\%)             \\ \cline{2-7}
                      & Chisqr   & 0.7553             & 0.0141             & 5.1490 (24.19\%)            & 5.5586 (21.83\%)            & 5.8526 (20.46\%)            \\ \cline{2-7}
                      & Log-Cosh & 0.9732             & \textbf{0.0074}    & \textit{6.5534 (3.51\%)}    & 6.7091 (5.65\%)             & 6.9818 (5.11\%)             \\ \cline{2-7}
\multirow{-4}{*}{Uni} & MSE+KL   & \textbf{0.9848}    & 0.0106             & \underline{6.5748 (3.20\%)} & 6.7207 (5.49\%)             & 6.9180 (5.98\%)             \\ \hline
                      & MSE      & 0.9532             & 0.0221             & 6.3528 (6.47\%)             & \textit{6.9148 (2.76\%)}    & 6.5403 (11.11\%)            \\ \cline{2-7}
                      & Chisqr   & 0.9385             & 0.0109             & 6.4832 (4.55\%)             & 6.5907 (7.32\%)             & \textit{7.0620 (4.02\%)}    \\ \cline{2-7}
                      & Log-Cosh & 0.9566             & 0.0129             & 6.4806 (4.59\%)             & 7.4687 (5.02\%)             & \underline{7.1165 (3.28\%)} \\ \cline{2-7}
\multirow{-4}{*}{IS}  & MSE+KL   & 0.9679             & 0.0318             & 6.3457 (6.57\%)             & 6.5471 (7.93\%)             & 6.9058 (6.14\%)             \\ \hline
                      & MSE      & 0.9743             & 0.0141             & \textbf{6.6264 (2.44\%)}    & 6.5845 (7.40\%)             & 6.9118 (6.06\%)             \\ \cline{2-7}
                      & Chisqr   & 0.9472             & \underline{0.0076} & 6.4752 (4.67\%)             & 6.7543 (5.01\%)             & 6.9806 (5.13\%)             \\ \cline{2-7}
                      & Log-Cosh & \underline{0.9826} & 0.0154             & 6.4898 (4.45\%)             & \underline{6.9491 (2.28\%)} & \textbf{7.1191 (3.24\%)}    \\ \cline{2-7}
\multirow{-4}{*}{HMC} & MSE+KL   & 0.9539             & 0.0171             & 6.3492 (6.52\%)             & \textbf{7.0110 (1.40\%)}    & 6.6109 (10.15\%)            \\ \hline\hline
\end{tabular}
\caption{Complete QNN results for the Breit-Wigner benchmark. 
Reported are $R^2$, $W_1$, and predicted integrals over the resonance intervals $[M_Z \pm 3\Gamma]$, $[M_Z \pm 5\Gamma]$, and $[M_Z \pm 10\Gamma]$. 
Parentheses indicate relative errors with respect to the analytic references $6.7924 \times 10^{-5}$, $7.1113 \times 10^{-5}$, and $7.3582 \times 10^{-5}$. 
All integral values are scaled by $10^{-5}$. 
These results extend Table~\ref{tab:BW_result} in the main text.
The values are emphasized with \textbf{bold} for the best, \underline{underline} for the second and \textit{italic} for the third value.}
\label{tab:qnn_bw_result}
\end{table*}

Finally, for the BW distribution for the Z boson Table~\ref{tab:qnn_bw_result} shows the full results expanding Table~\ref{tab:BW_result}.
QNN serves the $R^2$ scores remained at levels between $0.94 \sim 0.98$, while (Uniform, MSE+KL) records the highest $R^2$ score $0.9849$.
The (Uniform, Log-Cosh) combination shows the best integral behavior of the total training interval, $W_1 = 0.0074$.
Near to the resonance subintervals, (Uniform, MSE+KL) presents the best results in $[M_Z \pm 3\Gamma]$ with $2.44\%$, while for the wider intervals HMC sampling reveals better results as in $[M_Z \pm 5\Gamma]$ with MSE+KL loss has $1.40\%$ error and in $[M_Z \pm 10\Gamma]$ with Log-Cosh has $3.24\%$ error.
This indicates that for a singular structure like a reasonance peak, it is more optimal to use HMC sampling with Log-Cosh or MSE+KL loss. 
Especially, with $R^2$ score, we can use the (HMC, Log-Cosh).

Overall, the QNN results show reliable learning results with $R^2$ score over $0.9$ in QuInt-Net.
For the integral performance, we could consider the IS for the functions with an oscillatory structure or a discontinuous point.
However, it is more optimal to use the HMC witha  regularized loss function for the resonance-dominated case.

\subsection{QSP Results}\label{appendix:qsp}

Quantum Signal Processing (QSP) implements structured polynomial transformations~\cite{Martyn_2021}. 
In both $R^2$ and $W_1$ metrics, QSP's overall accuracy is typically lower than that of QNN. 

The QSP model is constructed with two operators: one is the signal operator:
\begin{equation}
S(x) = 
\begin{bmatrix} 
x & i\sqrt{1 - x^2} \\ 
i\sqrt{1 - x^2} & x 
\end{bmatrix},
\end{equation}
implemented via $\phi = 2\cos^{-1}(x)$ in an $RX(\phi)$ gate, and the other is the phase rotation
\begin{equation}
W(\theta) = e^{-i\theta \hat{\sigma}_Z/2}.
\end{equation}
The full unitary sequence
\begin{equation}
U_{\vec{\theta}}(x) = W(\theta_{d+1}) \prod_{i=1}^d S(x) W(\theta_i),
\end{equation}
implements a polynomial transformation whose real part is extracted using a Hadamard test,
\begin{equation}
\text{Poly}(x) = \text{Re}[P(x)] 
= \langle + | U_{\text{QSP}} | + \rangle.
\end{equation}
For a general target function $f(x)$, we decompose it into even and odd parts,
\begin{equation}
f(x) = 
\underbrace{\tfrac{f(x) + f(-x)}{2}}_{\text{Even}} 
+ \underbrace{\tfrac{f(x) - f(-x)}{2}}_{\text{Odd}},
\end{equation}
approximating them with degree-$d$ and degree-$(d-1)$ polynomials, respectively. 
This decomposition allows QSP to capture both symmetric and antisymmetric features within a compact circuit design~\ref{fig:Circuit}(b).

Table~\ref{tab:qsp_cpf_result}--\ref{tab:qsp_bw_result} summarize the QSP results for our benchmarking functions. 
While QSP circuits effectively capture local polynomial structures and reduce boundary artifacts, their limited expressivity compared to QNN restricts accuracy.

\begin{table*}[!h]
\setlength{\tabcolsep}{13pt}
\renewcommand{\arraystretch}{1.2}
\centering
\begin{tabular}{l|l|c|c|c|c|c}
\hline \hline 
Sampling              & Loss     & $R^2$              & $W_1$              & $[0, \pi/2]$                & $[-\pi/2, \pi/2]$             & $[-\pi, \pi]$       \\ \hline\hline
                      & MSE      & \underline{0.8798} & \textbf{0.0647}    & 0.8965 (15.98\%)            & 1.8207 (14.69\%)             & 0.0776              \\ \cline{2-7}
                      & Chisqr   & \textit{0.8797}    & 0.0923             & 0.8577 (19.61\%)            & 1.7221 (19.31\%)             & \textbf{-0.0158}    \\ \cline{2-7}
                      & Log-Cosh & \textbf{0.8822}    & \textit{0.0828}    & 0.9112 (14.60\%)            & 1.8570 (12.98\%)             & \textit{0.0292}     \\ \cline{2-7}
\multirow{-4}{*}{Uni} & MSE+KL   & 0.7847             & 0.3253             & 0.9166 (14.10\%)            & 1.8430 (13.64\%)             & 0.8629              \\ \hline
                      & MSE      & 0.8742             & 0.1158             & \underline{0.9986 (6.41\%)} & \underline{2.0008 (6.25\%)}  & -0.0581             \\ \cline{2-7}
                      & Chisqr   & 0.8765             & 0.1855             & 0.9536 (10.63\%)            & 1.9304 (9.55\%)              & -0.1202             \\ \cline{2-7}
                      & Log-Cosh & 0.8755             & \underline{0.0706} & \textit{0.9927 (6.96\%)}    & 1.9980 (6.38\%)              & \underline{-0.0185} \\ \cline{2-7}
\multirow{-4}{*}{IS}  & MSE+KL   & 0.8776             & 0.0951             & 0.9642 (9.64\%)             & 1.9543 (8.42\%)              & -0.0764             \\ \hline
                      & MSE      & 0.8340             & 0.4067             & \textbf{1.0274 (3.71\%)}    & \textbf{2.0887 (2.13\%)}     & 0.7716              \\ \cline{2-7}
                      & Chisqr   & 0.8260             & 0.4272             & 0.9481 (11.14\%)            & 1.9718 (7.60\%)              & 0.5810              \\ \cline{2-7}
                      & Log-Cosh & 0.8079             & 0.4291             & 1.0501 (1.59\%)             & 1.9947 (6.53\%)              & 0.3790              \\ \cline{2-7}
\multirow{-4}{*}{HMC} & MSE+KL   & 0.7903             & 0.4320             & 0.9860 (7.59\%)             & \textit{1.9975 (6.40\%)}     & 0.4508              \\ \hline\hline
\end{tabular}
\caption{Complete QSP results for the CPF. 
Metrics and interval definitions are identical to Table~\ref{tab:qnn_cpf_result}. 
Compared to QNN, QSP exhibits an underperformed $R^2$ score and $W_1$ distance.
The values are emphasized with \textbf{bold} for the best, \underline{underline} for the second, and \textit{italic} for the third value.}
\label{tab:qsp_cpf_result}
\end{table*}

The Table~\ref{tab:qsp_cpf_result} shows the CPF results.
Across the $R^2$ score, QSP has difficulty approximating the complex oscillatory structure, where it remained between $0.78 \sim 0.88$.
And it shows larger $W_1$ distances, which means it struggles with the global integral tendency.
The (HMC, MSE) presents a high accuracy in the subintervals, $[0, \pi/2], [-\pi/2, \pi/2]$, but it has a high inaccuracy in $[-\pi, \pi]$.

\begin{table*}[!h]
\setlength{\tabcolsep}{13pt}
\renewcommand{\arraystretch}{1.2}
\centering
\begin{tabular}{l|l|c|c|c|c|c}
\hline \hline 
Sampling              & Loss     & $R^2$              & $W_1$              & $[0, 0.5]$                  & $[-0.5, 0]$                  & $[-0.5, 0.5]$       \\ \hline\hline
                      & MSE      & \textbf{0.9393}    & 0.0340             & 0.2237 (10.52\%)            & -0.2230 (10.78\%)            & \underline{0.0006}  \\ \cline{2-7}
                      & Chisqr   & 0.9294             & 0.0339             & 0.2279 (8.84\%)             & -0.2207 (11.72\%)            & 0.0072              \\ \cline{2-7}
                      & Log-Cosh & 0.9300             & 0.0369             & 0.2213 (11.48\%)            & -0.2213 (11.48\%)            & \textbf{-0.00006}   \\ \cline{2-7}
\multirow{-4}{*}{Uni} & MSE+KL   & 0.9286             & 0.0373             & 0.2230 (10.76\%)            & -0.2138 (14.47\%)            & 0.0092              \\ \hline
                      & MSE      & 0.8956             & 0.0366             & 0.2687 (7.48\%)             & -0.2634 (5.36\%)             & 0.0053              \\ \cline{2-7}
                      & Chisqr   & 0.8969             & 0.0384             & \textit{0.2684 (7.36\%)}    & \textbf{-0.2562 (2.49\%)}    & 0.0122              \\ \cline{2-7}
                      & Log-Cosh & 0.8977             & \underline{0.0327} & \underline{0.2675 (7.00\%)} & \textit{-0.2632 (5.28\%)}    & 0.0042              \\ \cline{2-7}
\multirow{-4}{*}{IS}  & MSE+KL   & 0.9074             & \textbf{0.0251}    & \textbf{0.2545 (1.82\%)}    & \underline{-0.2628 (5.12\%)} & -0.0082             \\ \hline
                      & MSE      & \textbf{0.9393}    & 0.0365             & 0.2233 (10.68\%)            & -0.2212 (11.52\%)            & \textit{0.0021}     \\ \cline{2-7}
                      & Chisqr   & 0.9297             & 0.0379             & 0.2222 (11.11\%)            & -0.2162 (13.52\%)            & 0.0059              \\ \cline{2-7}
                      & Log-Cosh & \underline{0.9389} & 0.0338             & 0.2216 (11.35\%)            & -0.2174 (13.04\%)            & 0.0041              \\ \cline{2-7}
\multirow{-4}{*}{HMC} & MSE+KL   & \textit{0.9378}    & \textit{0.0329}    & 0.2263 (9.48\%)             & -0.2120 (15.19\%)            & 0.0142              \\ \hline\hline
\end{tabular}
\caption{Complete QSP results for the STEP function. 
Metrics and interval definitions follow Table~\ref{tab:qnn_step_result}. 
The values are emphasized with \textbf{bold} for the best, \underline{underline} for the second and \textit{italic} for the third value.}
\label{tab:qsp_step_result}
\end{table*}

For the Step function, the full results from QSP are shown in Table~\ref{tab:qsp_step_result}.
Notably, QSP achieves a reasonably high $R^2$ score ($0.89 \sim 0.93$) for this function, although it is less than the QNN model.
The global integral behavior shows a larger distance than QNN and shows unstable results in subinterval integrals.
However, (Uniform, Log-Cosh) achieves the most accurate integral result with $|\Delta I | = 0.00006$.

\begin{table*}[!h]
\setlength{\tabcolsep}{11pt}
\renewcommand{\arraystretch}{1.2}
\centering
\begin{tabular}{l|l|c|c|c|c|c}
\hline \hline 
Sampling              & Loss     & $R^2$              & $W_1$              & $[M_Z \pm 3\Gamma]$         & $[M_Z \pm 5\Gamma]$         & $[M_Z \pm 10\Gamma]$        \\ \hline\hline
                      & MSE      & \textbf{0.4851}    & 0.0564             & \underline{7.4155 (9.17\%)} & \underline{7.8174 (9.92\%)} & \textbf{7.4676 (1.48\%)}    \\ \cline{2-7}
                      & Chisqr   & 0.0662             & 0.0737             & 1.5090 (77.78\%)            & 2.0573 (71.06\%)            & 2.2330 (69.65\%)            \\ \cline{2-7}
                      & Log-Cosh & 0.3067             & 0.0617             & \textit{5.6599 (16.67\%)}   & \textbf{7.6702 (7.86\%)}    & \textit{6.8551 (6.83\%)}    \\ \cline{2-7}
\multirow{-4}{*}{Uni} & MSE+KL   & \underline{0.4151} & \textit{0.0438}    & \textbf{6.9821 (2.79\%)}    & \textit{8.1398 (14.46\%)}   & \underline{7.1696 (2.56\%)} \\ \hline
                      & MSE      & 0.1046             & 0.0466             & 11.4227(68.16\%)            & 10.5944 (48.98\%)           & 10.9515 (48.83\%)           \\ \cline{2-7}
                      & Chisqr   & \textit{0.3998}    & 0.0509             & 5.099 (24.91\%)             & 6.0790 (14.51\%)            & 6.0781 (17.39\%)            \\ \cline{2-7}
                      & Log-Cosh & 0.0573             & 0.0624             & 11.7395 (72.83\%)           & 12.0584 (69.56\%)           & 10,02512 (36.24\%)          \\ \cline{2-7}
\multirow{-4}{*}{IS}  & MSE+KL   & 0.0043             & 0.0660             & 11.4957 (69.24\%)           & 10.49550 (47.58\%)          & 9.5989 (30.45\%)            \\ \hline
                      & MSE      & 0.1630             & \textbf{0.0339}    & 12.1574 (78.98\%)           & 13.4724 (89.45\%)           & 12.42578 (68.86\%)          \\ \cline{2-7}
                      & Chisqr   & 0.3747             & \underline{0.0434} & 4.6532 (31.49\%)            & 5.6053 (21.17\%)            & 5.6542 (23.15\%)            \\ \cline{2-7}
                      & Log-Cosh & 0.2275             & 0.0602             & 12.1507 (78.88\%)           & 15.5869 (119.18\%)          & 12.46771 (69.43\%)          \\ \cline{2-7}
\multirow{-4}{*}{HMC} & MSE+KL   & 0.0735             & 0.0664             & 11.7572 (73.09\%)           & 10.8637 (52.76\%)           & 10.0829 (37.03\%)           \\ \hline\hline
\end{tabular}
\caption{Complete QSP results for the Breit-Wigner. 
Metrics and interval definitions follow Table~\ref{tab:qnn_bw_result}. 
The values are emphasized with \textbf{bold} for the best, \underline{underline} for the second, and \textit{italic} for the third value.}
\label{tab:qsp_bw_result}
\end{table*}

As shown in Table~\ref{tab:qsp_bw_result}, across all the combinations, the QSP model's $R^2$ scores dropped below 0.5, indicating that the model derivative fails to learn the function.
Comparing the subinterval results, there are some good integral results, but they lack consistency. 
For example, the (Uniform, MSE+KL) achieves low errors in $[M_Z \pm 3\Gamma]$ and $[M_Z \pm 10\Gamma]$, however, in $[M_Z \pm 5\Gamma]$, the error exceeds $10\%$.

In summary, the complete results of QSP shown in Table~\ref{tab:qsp_cpf_result}--\ref{tab:qsp_bw_result} indicate that the QSP model is not suitable to implement the QuInt-Net.
Specifically, QSP fails to learn the BW function, as all combinations yield $R^2$ scores below $0.5$.
For the CPF example, the $R^2$ score remained between $0.78$ and $0.88$, also exhibiting large $W_1$ distances.
While in the Step function, the QSP model serves relatively higher $R^2$ scores($0.89\sim0.93$), but still it underperforms compared to the QNN model and shows unstable limit for QuInt-Net.

\subsection{DQC1 Results}\label{appendix:dqc1}

The DQC1 model is evaluated as a hardware-efficient ansatz that requires only a single high-purity qubit while the remaining register can be maximally mixed~\cite{knill1998power}. 
This design with the NMR platform provides robustness against decoherence and allows for deeper circuit structures, making DQC1 a promising candidate for near-term noisy. 
Although DQC1's accuracy is still lower than that of QNN in both $R^2$ and $W_1$ metrics, it produces stable approximations of the integrand's smooth and global features in our study.

The DQC1 protocol estimates the normalized trace of a unitary operator---a classically intractable task---while requiring minimal coherence. 
The clean qubit is prepared with a Hadamard gate, followed by a controlled-unitary operation on the mixed-state register, yielding expectation values
\begin{equation}
\langle \sigma_X \rangle = \tfrac{\alpha}{2^n}\text{Re}[\text{Tr}(U)], 
\quad 
\langle \sigma_Y \rangle = \tfrac{\alpha}{2^n}\text{Im}[\text{Tr}(U)].
\end{equation}
The parameterized unitary is constructed as
\begin{equation}
U(x;\theta) = \prod_{l=1}^L W_l(\theta_l) V_l(x),
\end{equation}
where the data embedding layer is diagonal,
\begin{equation}
V_l(x) = \exp\!\left(-\tfrac{i}{2} x \sum_q \sigma_z^q \right).
\end{equation}
The trace then admits a Fourier-like decomposition,
\begin{equation}
\text{Tr}(U(x;\theta)) = \sum_{\omega \in \Omega} c_\omega(\theta) e^{i \omega x},
\end{equation}
enabling frequency-domain approximations of target functions with minimal quantum resources.

As shown in Fig.~\ref{fig:Circuit}, we construct the circuit with additional gates because of the property of the NMR system.
For the training parameter block $W(x)$, we use the three Pauli rotation gates, $W(x) = R_X(\theta_1) R_Y(\theta_2) R_Z(\theta_3)$, and then entangle the two qubits with a CZ gate.
Then, the embedding block is designed with a parameterized Pauli-$X$ gate, $S(x) = R_X(\theta \cdot x)$.
All of the gates are controlled with the clean qubit.
Then, Table~\ref{tab:dqc1_cpf_result}--\ref{tab:dqc1_bw_result} present the numerical results for CPF, STEP, and BW. 

\begin{table*}[!h]
\setlength{\tabcolsep}{13pt}
\renewcommand{\arraystretch}{1.2}
\centering
\begin{tabular}{l|l|c|c|c|c|c}
\hline \hline 
Sampling              & Loss     & $R^2$              & $W_1$              & $[0, \pi/2]$                & $[\pi/2, \pi/2]$             & $[-\pi, \pi]$       \\ \hline\hline
                      & MSE      & \underline{0.8792} & 0.0792             & 0.9284 (12.99\%)            & 1.8727 (12.25\%)             & \textbf{-0.0021}    \\ \cline{2-7}
                      & Chisqr   & 0.8783             & 0.0779             & 0.9499 (10.98\%)            & 1.9225 (9.91\%)              & 0.0523              \\ \cline{2-7}
                      & Log-Cosh & \textit{0.8790}    & \underline{0.0707} & 0.9230 (13.50\%)            & 1.8891 (11.48\%)             & \underline{-0.0184} \\ \cline{2-7}
\multirow{-4}{*}{Uni} & MSE+KL   & \textbf{0.8793}    & \textit{0.0717}    & 0.9284 (12.99\%)            & 1.8820 (11.81\%)             & -0.0416             \\ \hline
                      & MSE      & 0.8740             & 0.0896             & 0.9893 (7.29\%)             & 1.9921 (6.65\%)              & -0.1202             \\ \cline{2-7}
                      & Chisqr   & 0.8739             & \textbf{0.0667}    & 1.0241 (4.02\%)             & 2.0310 (4.83\%)              & \textit{-0.0194}    \\ \cline{2-7}
                      & Log-Cosh & 0.8728             & 0.0832             & 0.9691 (9.18\%)             & 1.9739 (7.51\%)              & -0.1500             \\ \cline{2-7}
\multirow{-4}{*}{IS}  & MSE+KL   & 0.8766             & 0.0932             & 1.0314 (3.34\%)             & 2.0507 (3.91\%)              & 0.0377              \\ \hline
                      & MSE      & 0.7813             & 0.3266             & \textbf{1.0743 (0.67\%)}    & \textbf{2.1497 (0.72\%)}     & 0.4947              \\ \cline{2-7}
                      & Chisqr   & 0.7954             & 0.2755             & \textit{1.0863 (1.80\%)}    & \underline{2.1007 (1.56\%)}  & 0.2834              \\ \cline{2-7}
                      & Log-Cosh & 0.7809             & 0.3159             & 0.9989 (6.38\%)             & 2.0626 (3.35\%)              & 0.2898              \\ \cline{2-7}
\multirow{-4}{*}{HMC} & MSE+KL   & 0.7730             & 0.2889             & \underline{1.0839 (1.57\%)} & \textit{2.1926 (2.73\%)}     & 0.4523              \\ \hline\hline
\end{tabular}
\caption{Complete DQC1 results for the CPF benchmark. 
Metrics and interval definitions follow Table~\ref{tab:qnn_cpf_result} 
The values are emphasized with \textbf{bold} for the best, \underline{underline} for the second and \textit{italic} for the third value.}
\label{tab:dqc1_cpf_result}
\end{table*}

The $R^2$ scores achieved by DQC1 are comparable to those of the QSP model.
It reaches only about $0.87 \sim 0.88$ for the uniform and IS, while HMC shows lower scores.
While QSP struggles with the $W_1$ distance, the DQC1 model has a more stable $W_1$ distance, indicating that it better captures the global integral behavior than the QSP model.
There are valuable insights in the subinterval integral, including the complex oscillatory structure in (HMC, MSE) with less than 1
However, it is not reliable because the low R² score suggests that the high accuracy observed in that interval is more likely due to chance than robust learning, indicating limitations in the model's overall predictive capability.

\begin{table*}[!h]
\setlength{\tabcolsep}{13pt}
\renewcommand{\arraystretch}{1.2}
\centering
\begin{tabular}{l|l|c|c|c|c|c}
\hline \hline 
Sampling              & Loss     & $R^2$              & $W_1$              & $[0, 0.5]$                  & $[-0.5, 0]$                  & $[-0.5, 0.5]$       \\ \hline\hline
                      & MSE      & \textbf{0.9856}    & 0.0099             & 0.2402 (3.90\%)             & -0.2426 (2.92\%)             & \textit{-0.0024}    \\ \cline{2-7}
                      & Chisqr   & \underline{0.9850} & 0.0110             & 0.2356 (5.72\%)             & \textit{-0.2467 (1.29\%)}    & -0.0110             \\ \cline{2-7}
                      & Log-Cosh & 0.9826             & 0.0122             & 0.2335 (6.58\%)             & -0.2464 (1.40\%)             & -0.0129             \\ \cline{2-7}
\multirow{-4}{*}{Uni} & MSE+KL   & \textit{0.9835}    & 0.0116             & 0.2409 (3.60\%)             & -0.2390 (4.39\%)             & \underline{0.0019}  \\ \hline
                      & MSE      & 0.9771             & \textbf{0.0057}    & 0.2445 (2.16\%)             & -0.2454 (1.83\%)             & \textbf{-0.0008}    \\ \cline{2-7}
                      & Chisqr   & 0.9830             & \underline{0.0063} & \textit{0.2469 (1.21\%)}    & -0.2461 (1.55\%)             & \textbf{0.0008}     \\ \cline{2-7}
                      & Log-Cosh & 0.9785             & \textit{0.0067}    & \underline{0.2505 (0.23\%)} & -0.2445 (2.16\%)             & 0.0059              \\ \cline{2-7}
\multirow{-4}{*}{IS}  & MSE+KL   & 0.9770             & 0.0088             & 0.2450 (1.98\%)             & \textbf{-0.2522 (0.88\%)}    & -0.0071             \\ \hline
                      & MSE      & 0.9830             & 0.0078             & 0.2382 (4.70\%)             & -0.2445 (2.16\%)             & -0.0063             \\ \cline{2-7}
                      & Chisqr   & 0.9825             & 0.0210             & 0.2339 (6.42\%)             & -0.2425 (2.96\%)             & -0.0086             \\ \cline{2-7}
                      & Log-Cosh & 0.9830             & 0.0110             & \textbf{0.2504 (0.19\%)}    & -0.2391 (4.33\%)             & 0.0113              \\ \cline{2-7}
\multirow{-4}{*}{HMC} & MSE+KL   & 0.9844             & 0.0115             & 0.2417 (3.30\%)             & \underline{-0.2471 (1.14\%)} & -0.0053             \\ \hline\hline
\end{tabular}
\caption{Complete DQC1 results for the STEP function. 
Metrics and interval definitions follow Table~\ref{tab:qnn_step_result}. 
The values are emphasized with \textbf{bold} for the best, \underline{underline} for the second and \textit{italic} for the third value.}
\label{tab:dqc1_step_result}
\end{table*}

The Table~\ref{tab:dqc1_step_result} shows that DQC1 performs comparably to QNN in the Step function example. 
The DQC1 model achieves a high $R^2$ score ($>0.97$), and while it records a slightly larger $W_1$ distance.
However, the DQC1 shows better results than QNN for the sampling method, IS.
Notably, (IS, $\chi^2$ Loss) achieves balance with $R^2=0.9830$, $W_1 = 0.0063$, and the relative errors below $2\%$ across subintervals including the discontinuity.
The results indicate that the DQC1 model can handle step-like discontinuities effectively under QuInt-Net, though its reduced expressivity.

\begin{table*}[!h]
\setlength{\tabcolsep}{11pt}
\renewcommand{\arraystretch}{1.2}
\centering
\begin{tabular}{l|l|c|c|c|c|c}
\hline \hline 
Sampling              & Loss     & $R^2$              & $W_1$              & $[M_Z \pm 3\Gamma]$         & $[M_Z \pm 5\Gamma]$         & $[M_Z \pm 10\Gamma]$        \\ \hline\hline
                      & MSE      & \textit{0.9405}    & 0.0322             & \textbf{6.8824 (1.32\%)}    & \textbf{7.2464 (1.89\%)}    & \textbf{7.4359 (1.05\%)}    \\ \cline{2-7}
                      & Chisqr   & 0.3013             & 0.0368             & 3.1399 (53.77\%)            & 3.4887 (50.94\%)            & 3.6247 (50.73\%)            \\ \cline{2-7}
                      & Log-Cosh & 0.9323             & 0.0226             & 6.3217 (6.92\%)             & 6.6413 (6.60\%)             & 6.9571 (5.45\%)             \\ \cline{2-7}
\multirow{-4}{*}{Uni} & MSE+KL   & 0.9391             & 0.0256             & 6.2855 (7.46\%)             & 6.5414 (8.01\%)             & 7.1426 (2.92\%)             \\ \hline
                      & MSE      & 0.9318             & 0.0185             & \textit{6.4551 (4.96\%)}    & 6.5468 (7.93\%)             & 7.0287 (4.47\%)             \\ \cline{2-7}
                      & Chisqr   & 0.8509             & 0.0176             & 5.9728 (12.06\%)            & 6.4704 (9.01\%)             & 6.5322 (11.22\%)            \\ \cline{2-7}
                      & Log-Cosh & \textit{0.9405}    & 0.0298             & \underline{6.5408 (3.70\%)} & 6.6626 (6.30\%)             & 6.9583 (5.43\%)             \\ \cline{2-7}
\multirow{-4}{*}{IS}  & MSE+KL   & 0.9319             & 0.0391             & 6.1971 (8.76\%)             & \underline{6.8371 (3.85\%)} & \textit{7.2439 (1.55\%)}    \\ \hline
                      & MSE      & 0.9304             & \textit{0.0162}    & 6.1722 (9.13\%)             & 6.3961 (10.05\%)            & 6.8217 (7.29\%)             \\ \cline{2-7}
                      & Chisqr   & 0.7928             & 0.0211             & 6.4145 (5.56\%)             & \textit{6.8238 (4.04\%)}    & 6.6850 (9.14\%)             \\ \cline{2-7}
                      & Log-Cosh & \underline{0.9491} & \textbf{0.0100}    & 6.4418 (5.16\%)             & 6.6551 (6.41\%)             & \underline{7.2509 (1.45\%)} \\ \cline{2-7}
\multirow{-4}{*}{HMC} & MSE+KL   & \textbf{0.9578}    & \underline{0.0122} & 5.9613 (12.23\%)            & 6.4280 (9.60\%)             & 6.9741 (5.21\%)             \\ \hline\hline
\end{tabular}
\caption{Complete DQC1 results for the Breit-Wigner benchmark. 
Metrics and interval definitions follow Table~\ref{tab:qnn_bw_result}. 
The values are emphasized with \textbf{bold} for the best, \underline{underline} for the second, and \textit{italic} for the third value.}
\label{tab:dqc1_bw_result}
\end{table*}

Most combinations achieved an $R^2$ score to level $0.9$, demonstrating a reliable level of learning, but still lower than the QNN model.
The most reliable learning strategy, (HMC, MSE+KL) predicts the global integral trend accurately with $W_1 = 0.0122$.
However, in the subintervals integral calculation, it produces inaccurate results with error rates exceeding $10\%$.
While in the general learning, (Uniform, MSE) serves the third reliable $R^2$ score and achieves the lowest error in the subinterval integrals.
This suggests that in complex distributions such as the BW distribution, the DQC1 with the highest overall reliability does not necessarily accurately predict the value within a specific critical interval.(see Table~\ref{tab:dqc1_bw_result}

The DQC1 model reflects a trade-off between performance and practicality. 
For functions with singular structures, the DQC1  could be competitive under QuInt-Net compared to QNN, although it is still less precise than QNN. 
However, when the NMR system ensures deep circuits, the DQC1 model for QuInt-Net provides a compact and noise-tolerant framework, enabling actual implementation.

\clearpage

\bibliography{ref}

\end{document}